\documentclass[acmsmall]{acmart}

\acmJournal{TAAS}

\usepackage{amsmath,amsfonts}
\usepackage{esvect}
\usepackage{graphicx}
\usepackage{subfigure}
\usepackage{booktabs}
\usepackage{tabularx}
\usepackage{multirow}
\usepackage{balance}
\usepackage{xcolor}
\usepackage{enumitem}
\usepackage{mathtools}
\usepackage{wrapfig}
\usepackage{ragged2e}
\usepackage[all]{nowidow}
\usepackage{stfloats}
\usepackage{anyfontsize}
\usepackage{algorithm}

\usepackage{color}
\usepackage{colortbl}

\usepackage{lineno}
\usepackage{bbding}
\usepackage{ulem}
\usepackage{syntax}

\newcolumntype{Y}{>{\centering\arraybackslash}X}

\usepackage{colortbl}
\usepackage{listings}

\theoremstyle{definition}

\newtheorem{dfn}{Definition}[section]

\definecolor{javared}{rgb}{0.6,0,0} 
\definecolor{javagreen}{rgb}{0.25,0.5,0.35} 
\definecolor{javapurple}{rgb}{0.5,0,0.35} 
\definecolor{javadocblue}{rgb}{0.25,0.35,0.75} 
\definecolor{javagrey}{rgb}{0.46,0.45,0.48} 

\newcommand\resq[1]{
\noindent 
\fcolorbox{green!40!black}{green!5}{\noindent 
 \parbox{0.98\columnwidth}{\noindent  #1}}\\
}

\lstdefinestyle{Alg}{
  basicstyle=\ttfamily\footnotesize,
  breaklines=true,
  tabsize=2,
  mathescape,
  numbers=left,
  xleftmargin=2.5em,
  xrightmargin=0.5em,
  frame=tb,
  framexleftmargin=2em,
  emph={Algorithm,Input,Output,for,each,do,if,else,Function,while,let,be,repeat,until,return,times,and,or,break,in,then,},
  emphstyle={\textbf},
  escapechar=?,
  morecomment=[l][\color{javagreen}]{//},
  columns=flexible,
}

\usepackage{framed}
\usepackage{tikz}
\definecolor{light-gray}{gray}{0.9}
\usepackage[framemethod=TikZ]{mdframed}
\mdfsetup{skipabove=5pt,skipbelow=3pt}
\usepackage{lipsum}
\mdfdefinestyle{RQFrame}{%
	linecolor=black,
	outerlinewidth=0.15pt,
	roundcorner=3pt,
	innertopmargin=2pt,
	innerbottommargin=2pt,
	innerrightmargin=4pt,
	innerleftmargin=4pt,
	backgroundcolor=light-gray}

\title{Using Genetic Programming to Build Self-Adaptivity into Software-Defined Networks}

\author{Jia Li, Shiva Nejati and  Mehrdad Sabetzadeh}
\affiliation{%
  \institution{University of Ottawa, Canada}
}
\email{{jli714,snejati,m.sabetzadeh}@uottawa.ca}

\begin{document}

\renewcommand{\emph}[1]{\textit{#1}}


\begin{abstract}

Self-adaptation solutions need to periodically monitor, reason about, and adapt a running system. The adaptation step involves generating an adaptation strategy and applying it to the running system whenever an anomaly arises. In this article, we argue that, rather than generating individual adaptation strategies, the goal should be to adapt the control logic of the running system in such a way that the system itself would learn how to steer clear of future anomalies, without triggering self-adaptation too frequently. While the need for adaptation is never eliminated, especially noting the uncertain and evolving environment of complex systems, reducing the frequency of adaptation interventions is advantageous for various reasons, e.g., to increase performance and to make a running system more robust.

We instantiate and empirically examine the above idea for software-defined networking -- a key enabling technology for modern data centres and Internet of Things applications. Using \hbox{genetic programming\,(GP)}, we propose a self-adaptation solution that continuously learns and updates the control constructs in the data-forwarding logic of a software-defined network. Our evaluation, performed using open-source synthetic and industrial data, indicates that, compared to a baseline adaptation technique that attempts to generate individual adaptations, our GP-based approach is more effective in resolving network congestion, and further, reduces the frequency of adaptation interventions over time. In addition, we show that, for networks with the same topology, reusing over larger networks the knowledge that is learned on smaller networks leads to significant improvements in the performance of our GP-based adaptation approach. Finally, we compare our approach against a standard  data-forwarding algorithm from the network literature, demonstrating that our approach significantly reduces packet loss. 
\end{abstract}

\ccsdesc[500]{Software and its engineering~Search-based software engineering}
\ccsdesc[500]{Software and its engineering~Empirical software validation}
\ccsdesc[500]{Computing methodologies~Machine learning algorithms}

\keywords{Self-adaptive Systems, Search-based Software Engineering, Genetic Programming and Software-defined Networks.}


\maketitle


\section{Introduction}
\label{sec:intro}

A major challenge when engineering complex systems is to ensure that these systems  meet their quality-of-service criteria and are reliable in the face of uncertainty. Self-adaptation is a promising approach for addressing this challenge. 
The idea behind self-adaptation is that engineers take an existing system, specify its expected qualities and objectives as well as strategies to achieve these objectives, and build capabilities into the system in a way that enables the system to adjust itself to changes during operation~\cite{DBLP:books/daglib/p/GarlanSC09}. 

Many software-intensive systems can benefit from self-adaptivity. A particularly pertinent domain where self-adaptation is useful is software-defined networking -- a flexible network architecture that is prevalent in modern data centres and Internet of Things applications~\cite{Drutskoy:13, Wang:17, Rafique:20, ShinNSB0Z20}.
A software-defined network (SDN) provides centralized programmable control over distributed network resources, thereby allowing network operators to better manage network performance and to react in real time to events in the network.
%
%
Notably, many networks are prone to \emph{congestion} when there is a burst in demand. For instance, in an emergency management system, such bursts may occur when a disaster situation, e.g., a flood, is unfolding. 
Taking advantage of the software programmability of SDN controllers, our ultimate goal in this article is to develop a self-adaptive approach for resolving network congestion.

Self-adaptation has been studied for many years~\cite{Moreno:15,PaucarB19,JahanRWGPMC20,ChengRM13,DBLP:journals/taas/SalehieT09}. The existing self-adaptation approaches include  model-based~\cite{DBLP:conf/models/DeVriesC17,Ramirez:10}, control-based~\cite{Filieri:15}, requirements-based~\cite{DBLP:conf/icse/AlrajehCL20}, and more recently, learning-based~\cite{DBLP:conf/seams/GheibiWQ21} solutions. At the heart of all self-adaptation solutions, there is a planning step that generates or determines an adaptation strategy to adapt the running system once an anomaly is detected~\cite{DBLP:books/daglib/p/GarlanSC09}. Adaptation strategies may be composed of fixed and pre-defined actions identified based on domain knowledge, or they may be new behaviours or entities introduced at run-time~\cite{DBLP:journals/taas/SalehieT09}. Irrespective of the type of adaptation, the knowledge about how to generate adaptation strategies is often concentrated in the planner: The running system is merely modified by the planner to be able to handle an anomaly as seen in a specific time and context; the running system is not necessarily improved in a way that it can better respond to future anomalies on its own. As we illustrate in our motivating example of Section~\ref{sec:motivate}, having to repeatedly invoke the planner for each adaptation can be both expensive and ineffective. 
The main idea that we put forward in this article is as follows: \emph{The self-adaptation planner should attempt to improve the control logic of the running system such that the system itself would learn how to steer clear of future anomalies,  without triggering self-adaptation too frequently.} The need for self-adaptation is never entirely eliminated, especially noting that the environment changes over time. Nonetheless, a system augmented with such a self-adaptation planner is likely to be more efficient, more robust to changes in the environment (e.g., varying network requests), and less in \hbox{need of adaptation intervention.}
Furthermore, existing self-adaptation research focuses on modifying a running system via producing individual and concrete elements, e.g., configuration values~\cite{Iftikhar:17}. In contrast, we take a  \emph{generative} approach~\cite{FeldtY20}, modifying the logic of the running system which in turn generates the concrete elements.

\textbf{Contributions.} We propose a generative self-adaptation framework, named GenAdapt, that uses \emph{genetic programming (GP)}~\cite{koza1992genetic,poli2008field} to realize the above-described vision of self-adaptation. GenAdapt builds on the well-known MAPE-K loop~\cite{kephart:03,Moreno:15} to incrementally enhance the control logic of the running systems. The idea is that by incremental adaptations (i.e., evolving a running system's logic), the running system becomes more robust to changes in the environment and requires less frequent adaptation. GenAdapt requires as input  a context-free grammar capturing the language in which the configurable (adaptive) fragment  of a system's control logic has been expressed. It then employs genetic programming to incrementally evolve the system's control logic so that the system responds better to environment changes observed over time. To ensure the continuity of learning over time and to evolve the control logic in a way that better fits the environment in which the system is operating, at each round of self-adaptation, we maintain in the MAPE-K knowledge base a set of best solutions (alternative control-logic implementations) computed by GP. These solutions are used to partially bootstrap GP in the future rounds of self-adaptation, thereby informing these future rounds about the candidate solutions that best fit the environment uncertainty in the past.

In this article, we  instantiate GenAdapt to address congestion control for SDNs. 
In an SDN,  network control is transferred from local fixed-behaviour controllers distributed over a set of switches to a centralized and programmable software controller~\cite{SDN:15}. More specifically, we address the self-adaptation of SDN data-forwarding algorithms in order to resolve network congestion in real time. We implement  GenAdapt  to enhance the  programmable controller of an SDN. Noting that GenAdapt relies on the MAPE-K loop, it performs the following tasks periodically:  (i)~monitoring the SDN to check if it is congested and generating a model of the SDN  to be used for congestion resolution; (ii)~applying GP to evolve the logic of the SDN data-forwarding algorithm such that congestion is resolved and, further, the transmission delay and the changes introduced in the existing  data-transmission routes  are minimized; and (iii)~modifying  existing transmission routes to resolve the current congestion and updating the logic of SDN data forwarding to mitigate future congestion. Exploiting the global network view provided by an SDN for modifying the controller and resolving congestion in real time is not new. However, existing approaches  adapt  the network management logic using  \emph{pre-defined} rules~\cite{Yashar:12}. In contrast, our approach uses GP to modify the data-forwarding control constructs in an evolving manner and without reliance on fixed rules.

We have implemented GenAdapt into a tool, which we make publicly available~\cite{appedix}. As we describe in more detail in Section~\ref{sec:tool}, the GenAdapt tool interacts with (i)~the (software-programmable) controller of an SDN, (ii)~a network emulator that simulates the network infrastructure, and (iii) a traffic generator that emulates different types of requests from devices and network users.

\textbf{Evaluation.} We evaluate GenAdapt on $26$ synthetic and one industrial IoT network. The industrial network, which is the backbone of an IoT-based emergency management system~\cite{ShinNSB0Z20}, is prone to congestion when the volume of demand increases during emergencies. We compare our framework with two baselines: (1) a self-adaptive technique from the SEAMS community that attempts to resolve congestion by generating individual adaptations (i.e., individual data-transmission routes) without optimizing the SDN routing algorithms~\cite{ShinNSB0Z20}; and (2)~a standard data-forwarding algorithm from the network community that uses pre-defined rules (heuristics) to optimize SDN control at runtime and resolve congestion~\cite{Coltun:08,Cisco:05}.

{Our results indicate that GenAdapt successfully resolves congestion in all the 26 networks considered while the adaptive, but non-generative baseline fails to resolve congestion in four of the networks with probabilities ranging from $10$\% to $66$\%. In addition, compared to this baseline, GenAdapt reduces the average number of congestion occurrences, and hence, the number of adaptation rounds necessary. Compared to the baseline from the network community, GenAdapt is able to significantly reduce packet loss for the industrial network.

In addition, we empirically assess the impact of transferring to larger networks the best control logic learned on smaller networks, when both the smaller and larger networks have the same topology (e.g., both are full graphs). We observe that for a given topology, bootstrapping GenAdapt with the best logic learned on a smaller network can significantly improve performance on a larger network in terms of the number of adaptation invocations, the amount of packet loss and overall duration the network remains congested. These results suggest the transferability of the learned logic over networks that share the same topology but have different numbers of nodes. 
}Finally, we empirically demonstrate that the execution time of GenAdapt is  linear in network size and the amount of data traffic over time, thus making it suitable for online adaptation.

\textbf{Comparison with conference version.} This article extends a previous conference paper~\cite{Li:22}, published at the 17th Symposium on Software Engineering for Adaptive and Self-Managing Systems (SEAMS~2022). In comparison to this earlier publication, this current article provides the following major extensions: (1)~We provide a high-level description of the GenAdapt framework to convey -- as independently as possible from our application context (i.e., SDN data forwarding) -- the main characteristics of the framework; (2) We considerably expand our earlier empirical evaluation in terms of the number of synthetic networks considered, going from 18 networks to 26 networks, and increasing the number of experiments by 50\%.   Extending our evaluation allows us to provide additional insights on improving effectiveness of our framework by  reusing adaptation solutions across networks;
(3) We empirically examine the bootstrapping of our framework and the impact of applying over a larger network the control logic that has been learned on a smaller network; 
(4) We provide a detailed description of our tool support; and, (5) We provide a more thorough discussion of related work and threats to validity.

\textbf{Structure.} Section~\ref{sec:motivate} motivates the paper. Section~\ref{sec:overview} provides an overview of our self-adaptation framework, GenAdapt. Section~\ref{sec:selfcontrol} presents an instantiation of GenAdapt in the context of SDN data forwarding. Section~\ref{sec:tool} presents our tool support.  Section~\ref{sec:eval} describes our evaluation. Section~\ref{sec:related} compares with related work. Section~\ref{sec:conclusions} concludes the paper. 
\section{Motivating Example}
\label{sec:motivate}
We motivate our approach through an example. Network-traffic management is about assigning flow paths to network requests such that the entire network is optimally utilized~\cite{Noormohammadpour18,Alizadeh:10,Agarwal:13,Betzler:16}. A flow path is a directed path of network links.  Many network systems use a standard, lightweight data-forwarding algorithm, known as Open Shortest Path First (OSPF)~\cite{Coltun:08,Cisco:05}. OSPF generates a flow path for a data-transmission request by computing the shortest weighted path between the source and destination nodes of the request.  

\begin{figure}[t]
	\centerline{\includegraphics[width=0.7\textwidth]{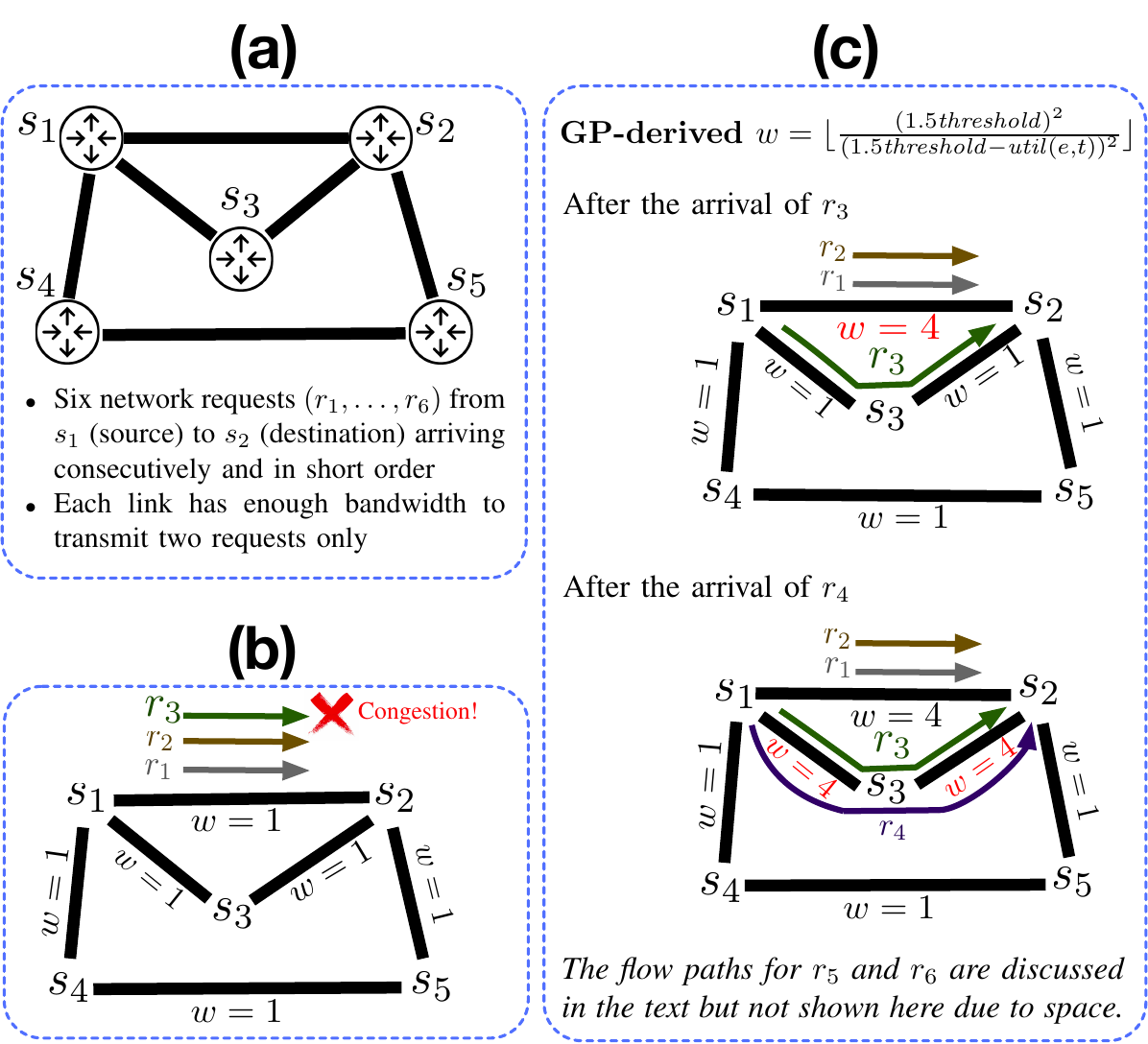}}
	\caption{(a) A simple network alongside the description of network requests; (b) Output of a baseline data-forwarding algorithm; and (c) Output of data-forwarding after adaptation by our approach using genetic programming.}
	\label{fig:example}
\end{figure}

Figure~\ref{fig:example}(a) shows a network with five nodes (switches) and six links. Suppose that there are six  requests, $r_1$, \ldots, $r_6$ from source node $s_1$ to destination node $s_2$, and that they arrive in sequence (first $r_1$, then $r_2$, \ldots) with a few seconds in between. These requests keep running after arrival. Further, all link weights are set to one. OSPF creates the flow path $s_1 \rightarrow s_2$ for every request.
Assume that each flow utilizes $30$\% of the bandwidth of each link. Further, assume that we have a threshold of $80$\% for link utilization, above which we consider a link to be congested. 
Each link in our example then has enough bandwidth to transmit two flows before it is considered congested. Consequently, link $s_1 \rightarrow s_2$ will be congested after the arrival of $r_3$; see Figure~\ref{fig:example}(b). This leads to an invocation of
the self-adaptation planning step to resolve the congestion. A common approach for congestion resolution is through combinatorial optimization, where some flow paths are re-routed such that the data stream passing through each link remains below the utilization threshold~\cite{FortzT02,DBLP:conf/cscwd/LiuZGQ18,6980429}. This optimization can  be done using various methods, e.g., graph optimization~\cite{FortzT02}, mixed-integer linear programming~\cite{Agarwal:13}, \hbox{local search~\cite{FortzT00,FortzT04}, and genetic algorithms~\cite{ShinNSB0Z20}.}

Optimization by re-routing flow paths has a major drawback. While optimizing at the level of individual flow paths may solve the currently observed  congestion, doing so does not improve the logic of OSPF and thus does not contribute to congestion avoidance in the future. For instance, in Figure~\ref{fig:example}(b), the arrival of $r_3$ will cause congestion. Re-routing $r_3$ over $s_1 \rightarrow s_3 \rightarrow s_2$ will resolve this congestion. But, upon the arrival of $r_4$, OSPF will yet again select the flow path $s_1 \rightarrow s_2$ for $r_4$ (since this is the shortest path between $s_1$ and $s_2$) just to find the $s_1 \rightarrow s_2$ link congested again. This in turn necessitates the self-adaptation planning step to be invoked for $r_4$ as well. In a similar vein, the arrival of $r_5$ and $r_6$ will cause congestion, prompting further calls to self-adaptation planning.

Our proposal is that, at each round of self-adaptation where a congestion is detected, one should focus on optimizing the logic of OSPF instead of optimizing flow paths. In this paper, we optimize and update the link-weight formula that OSPF uses. The hypothesis here is that an optimized link-weight formula not only can resolve the existing congestion, but can simultaneously also make OSPF more intelligent towards congestion avoidance in the future, thus reducing the number of times that the self-adaptation planning step has to be invoked.
Below, we illustrate our approach using the example of Figure~\ref{fig:example}. 

We use genetic programming (GP) for self-adaptation planning, whereby we dynamically learn link-weight functions that not only help resolve an existing congestion but also help steer clear of future ones.
Initially, and for requests $r_1$ and $r_2$, our approach does exactly as the standard OSPF would do, since there is no congestion. Upon the arrival of $r_3$ and the detection of congestion, i.e., the situation in Figure~\ref{fig:example}(b), our self-adaptation approach kicks in and automatically computes a link-weight formula to reroute some of the requests in order to resolve the congestion. For example, one possible formula that our approach could generate is  : $ \frac{(1.5\mathit{threshold)^2}}{(1.5\mathit{threshold} - \mathit{util}(e, t))^2}$. This formula would be stored in memory as a parse tree  and called to compute the  weight for each link when a new request arrives.  
In this formula,  $\mathit{util}(e, t)$ is the utilization percentage of link $e$ at time $t$, and  $\mathit{threshold}$ is a constant parameter describing the utilization threshold above which the network is considered to be congested. The weight values generated for the links are first taken as absolute values and then rounded down to integers. This ensures that only positive integer weights are assigned to the links. In our example, $\mathit{threshold}=0.8$, and for each link, $\mathit{util}(e, t)$ is 0.3 if one flow passes through $e$, and is 0.6 if two flows pass through $e$.  Given that  $r_1$ and $r_2$ are already routed through $s_1 \rightarrow s_2$, the weight of \hbox{$s_1 \rightarrow s_2$} computed by this formula becomes $w=4$; see Figure~\ref{fig:example}(c) after $r_3$'s arrival. As a result, OSPF selects path $s_1 \rightarrow s_3 \rightarrow s_2$ for $r_3$ which successfully resolves the current congestion observed in Figure~\ref{fig:example}(b). 
This, however, does not increase the weights for $s_1 \rightarrow s_3$ and $s_3 \rightarrow s_2$, since these links are utilized at $30$\%, and  their weights remain at 1. Once $r_4$ arrives, OSPF directs $r_4$ to $s_1 \rightarrow s_3 \rightarrow s_2$ as it is the shortest weighted path between $s_1$ and $s_4$. This will not cause any congestion, but the weights for $s_1 \rightarrow s_3$ and $s_3 \rightarrow s_2$ increase to $w=4$ since these links are now utilized at $60$\%. Finally, OSPF will direct the last two requests $r_5$ and $r_6$ to the longer path $s_1 \rightarrow s_4 \rightarrow s_5 \rightarrow s_2$, since, now, this path is the shortest \hbox{weighted path between $s_1$ and $s_2$.}

As shown above, using our GP-learned link-weight formula, OSPF is now able to manage flow paths without causing any congestion and without having to re-invoke the self-adaptation planning step beyond the single invocation after the arrival of $r_3$.

We note that there are several techniques that modify network parameters at runtime to resolve congestion in SDN (e.g.,~\cite{Yashar:12,PriyadarsiniMBK19}). These techniques, however, rely on pre-defined rules. For example, to make OSPF  -- discussed above -- adaptive, a typical approach is to  define a network-weight function~\cite{RetvariNCS09}. This function can involve parameters that change at runtime and based on the state of the network. However, the structure of the function is fixed. Consequently, the function may not be suitable for all networks with different characteristics and inputs. To address this limitation, we use GP to \emph{learn and evolve the function structure} instead of relying on a fixed, manually crafted function. Our approach is generative and seeks to identify a higher-order structure, specifically a link-weight formula, that results in optimal adaptation solutions. The generated link-weight formula is used to determine the flow paths for incoming requests. In Section~\ref{sec:eval}, we compare our GP-based approach with OSPF configured using an optimized weight function suggested by Cisco standards~\cite{Coltun:08,Cisco:05}. As we show there, OSPF's optimized weight function cannot address the congestion caused by the network-request bursts in our industrial case study.

In this paper, we focus on situations where congestion can be resolved by re-routing -- in other words, when the network topology is such that alternative flow paths can be created for some requests. Given our focus, OSPF is a natural comparison baseline, noting that, in OSPF, congestion is resolved by re-routing. When alternative flow paths do not exist, congestion resolution can be addressed only via traffic shaping~\cite{Yashar:12}, i.e., via modifying the network traffic. Our approach does not alter the network traffic. We therefore do not compare against congestion-resolution baselines \hbox{that use traffic shaping.}

\section{Framework Overview}
\label{sec:overview}
Figure~\ref{fig:fmw} shows an overview of GenAdapt -- our generative self-adaptation framework. The main goal of GenAdapt is to adapt a running system by dynamically improving its control logic (e.g., a formula or a logical condition based on which the system makes a decision).  GenAdapt aims to improve the designated logic in response to environmental changes and find a trade-off between self-adaptation objectives.  For example, in the context of SDN data forwarding, the control logic that GenAdapt is improving is the link-weight function used by the SDN's data-forwarding algorithm, as illustrated in Section~\ref{sec:motivate}. At each round of self-adaptation, GenAdapt evolves the link-weight function with the goal of satisfying a trade-off between three objectives related to the network's quality of service; these objectives will be defined and formalized in Section~\ref{subsec:selfloop}. 

GenAdapt leverages MAPE-K~\cite{kephart:03,Moreno:15}-- the well-known self-adaptation control-loop. As shown in Figure~\ref{fig:fmw}, MAPE-K has four main steps: 

\begin{figure}[t]
  \centerline{\includegraphics[width=.8\columnwidth]{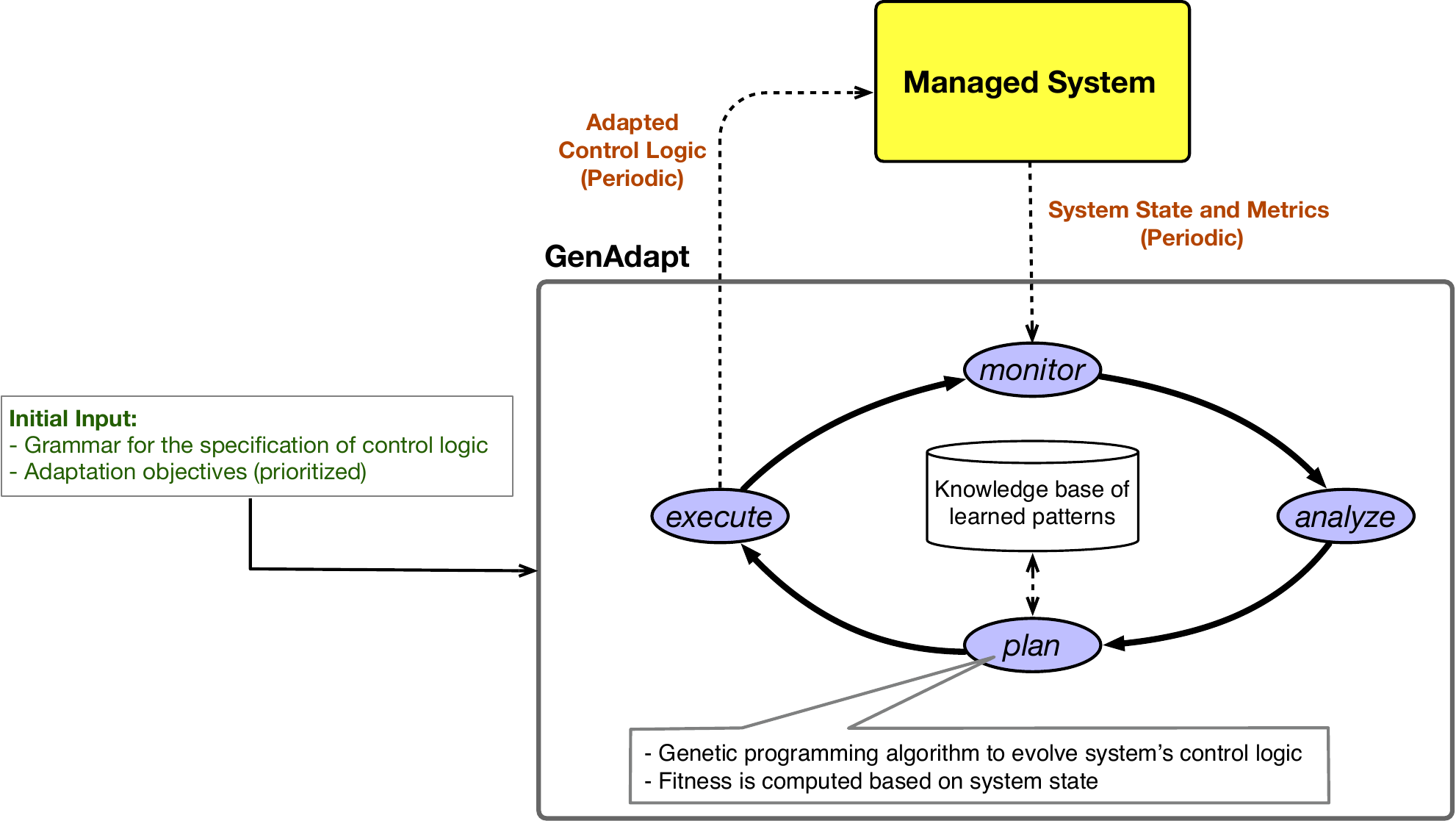}}
  \caption{Overview of the GenAdapt Framework.}
  \label{fig:fmw}
\end{figure}

\textbf{(1)} \emph{Monitoring} the system and its environment: At this step, GenAdapt periodically receives the following information from the running system: (I)~the \emph{state of the system} at the current time, and (II)~certain \emph{metrics}  that enable GenAdapt to monitor for the occurrence of anomalies (i.e., violations of self-adaptation goals).

\textbf{(2)} \emph{Analyzing} the information collected from the system and its environment and deciding whether adaptation is needed. 
At this stage, GenAdapt uses the system state and the metrics collected at the monitoring step to decide if some system goal has been violated, in which case the planning step needs to be triggered. 

\textbf{(3)} \emph{Planning} to adapt the system to address any violations observed during the previous step. This step employs a genetic programming (GP) algorithm to evolve the control logic of the running system such that when the system uses that evolved logic, the observed violation is no longer present, or its severity has been reduced. The GP algorithm employed in the planning step requires, in addition to the inputs received periodically from the running system, two further inputs: First, GP requires a context-free grammar specifying the language (template) for the control logic that is subject to modification. For example, we use a grammar defining valid mathematical formulas  consisting of basic arithmetic operations and network parameters to specify a template for candidate link-weight functions in SDN data forwarding. 
Second, GP requires a  set of prioritized adaptation objectives. The objectives should be defined so that optimizing them eliminates or reduces the violation that triggered adaptation. The GP algorithm aims to satisfy the objectives based on their order of priority. For objectives with the same priority, the planning step aims to find a trade-off solution that equally satisfies them. The grammar and the prioritized set of objectives should be provided at the outset and prior to launching GenAdapt.

\textbf{(4)} \emph{Executing} the adaptation by applying it to the system. This step applies to the running system the adaptation (i.e., the adapted control logic) computed by the planning step. The desired outcome here is that adaptation should not only address the current observed  anomaly but should further make the running system more robust so that the system itself can steer clear of anomalous situations that may arise again.

Our main focus is to improve the planning step of self-adaptation through the application of GP.  In designing our GP algorithm for the planning step, we consider two important factors:  

\emph{First}, for GP to assess the adaptation objectives for candidate solutions,  the system should be executed for each candidate. Since we cannot execute the system in the adaptation loop to compute objectives for every candidate generated during search, we use the system state collected from the running system at the monitoring step to compute the objectives. The underlying assumption here is that the main parameters of the system remain unchanged for the duration of running the self-adaptation loop. 

\emph{Second,} GenAdapt uses a knowledge base to store the best control logic computed at each round of the planning step. Best solutions generated by previous invocations of GenAdapt are reused in the initial population of GP whenever the planning step is re-invoked. Reusing as bootstrap some of the best solutions from the previous round ensures continuity and incrementality in learning adaptation solutions across multiple rounds of self-adaptation. Reusing the best solutions further aims to ensure that computed adaptations can effectively address uncertainties observed over a long period instead of being focused or over-fitted to the uncertainties observed in a short period.

For example, the formula in Figure~\ref{fig:example}(c) is generated as one of the best solutions at the adaptation round invoked upon the arrival of $r_3$ and is stored in the knowledge base. In future rounds of self-adaptation, we will use this formula to bootstrap the initial population of GP. Our conjecture is that this formula will likely solve future occurrences of congestion with minor mutations.  We test this conjecture in our empirical evaluation for the SDN domain (see RQ1 and RQ2 in Section~\ref{sec:eval}). We note that certain elements of GenAdapt might have SDN-specific characteristics, and it is important to recognize that our approach might not be applicable across all domains.

\section{Self-Adaptation for SDN}
\label{sec:selfcontrol}

\begin{figure}[t]
	\centerline{\includegraphics[width=.7\columnwidth]{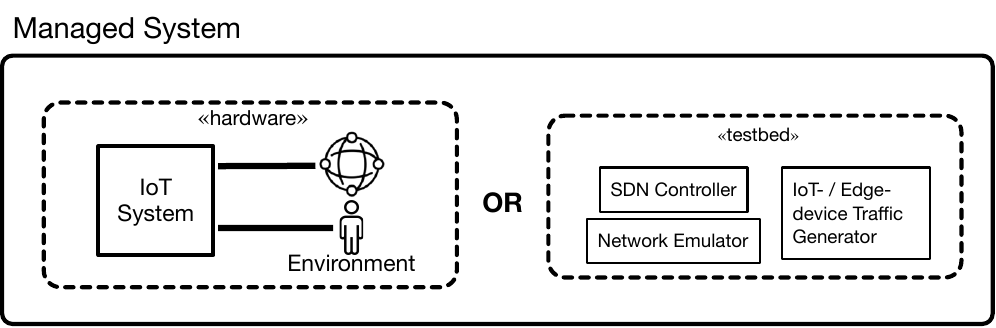}}
	\caption{The managed system interacting with GenAdapt may be a real or a testbed system.}\label{fig:adaptloop}
\end{figure}

We apply GenAdapt to make the data-forwarding logic of SDN controllers self-adaptive. For this, we implement GenAdapt as an add-on to the programmable SDN  control layer. In an SDN, the control layer has a global view of the entire network and can dynamically modify the network at runtime. The routing logic of an SDN is centralized, decoupled from data and network hardware, and expressed using software code at the control layer. Given the separation of control logic from data planes and forwarding hardware, changing link weights can be programmed at the control layer in a way that the changes do not affect the existing network flows and instead are used only for routing new flows~\cite{Apostolopoulos:99}. Dynamically modifying link weights therefore does not jeopardize network stability.

Figure~\ref{fig:adaptloop} shows a schematic view of an SDN (managed system) interacting with GenAdapt. The managed system can be either a system involving actual hardware, or a realistic testbed. In our work, GenAdapt interacts with a high-fidelity testbed that includes an actual, carrier-grade SDN controller and a simulated network. Using a simulated network rather than an actual network infrastructure is a common approach when designing and evaluating self-adaptation techniques. This enables us to experiment with a multitude of network systems rather than merely one fixed network  setup~\cite{paper2,paper5,Iftikhar:17,SotiropoulosWGI17}. 

As shown on the right side of Figure~\ref{fig:adaptloop}, our testbed combines three components: An SDN controller capturing the software-defined controller of a network system~\cite{Berde:14}; a network emulator that simulates the network  infrastructure including links, nodes and their properties~\cite{Lantz:10}; and a traffic generator~\cite{Botta:12} that emulates different types of requests generated by IoT devices and sensors.

To assess the fitness values, i.e., adaptation objectives, we cannot use the actual testbed and its environment for each candidate solution, i.e., each candidate link-weight function,  since this system is real-time (wall-clock-time) and requires a few milliseconds to compute the behaviour of a network for each candidate link-weight formula. Our GP algorithm, when invoked, needs to
explore a large number of candidate formulas. To do so efficiently, we use \emph{surrogate} computations that approximate the fitness values~\cite{paper1,paper3,paper4}. Specifically, as we detail in Section~\ref{subsec:selfloop}, GenAdapt computes the fitness for each candidate solution using a snapshot of the system and its environment from the testbed,  assuming that the system does not change during a small time period.  We set the maximum time budget for adaptation to one second, the smallest monitoring interval permitted by our SDN controller as we discuss in Section~\ref{sec:tool}. With regard to the general applicability of this assumption, we note that determining the time budget for self-adaptation requires careful consideration of the particular characteristics of the application domain; one needs to strike a balance where the allocated time is neither too short, impeding genetic improvement, nor too long, resulting in an outdated system state during the monitoring step.

\begin{figure}[t]
\centerline{\includegraphics[width=.65\columnwidth]{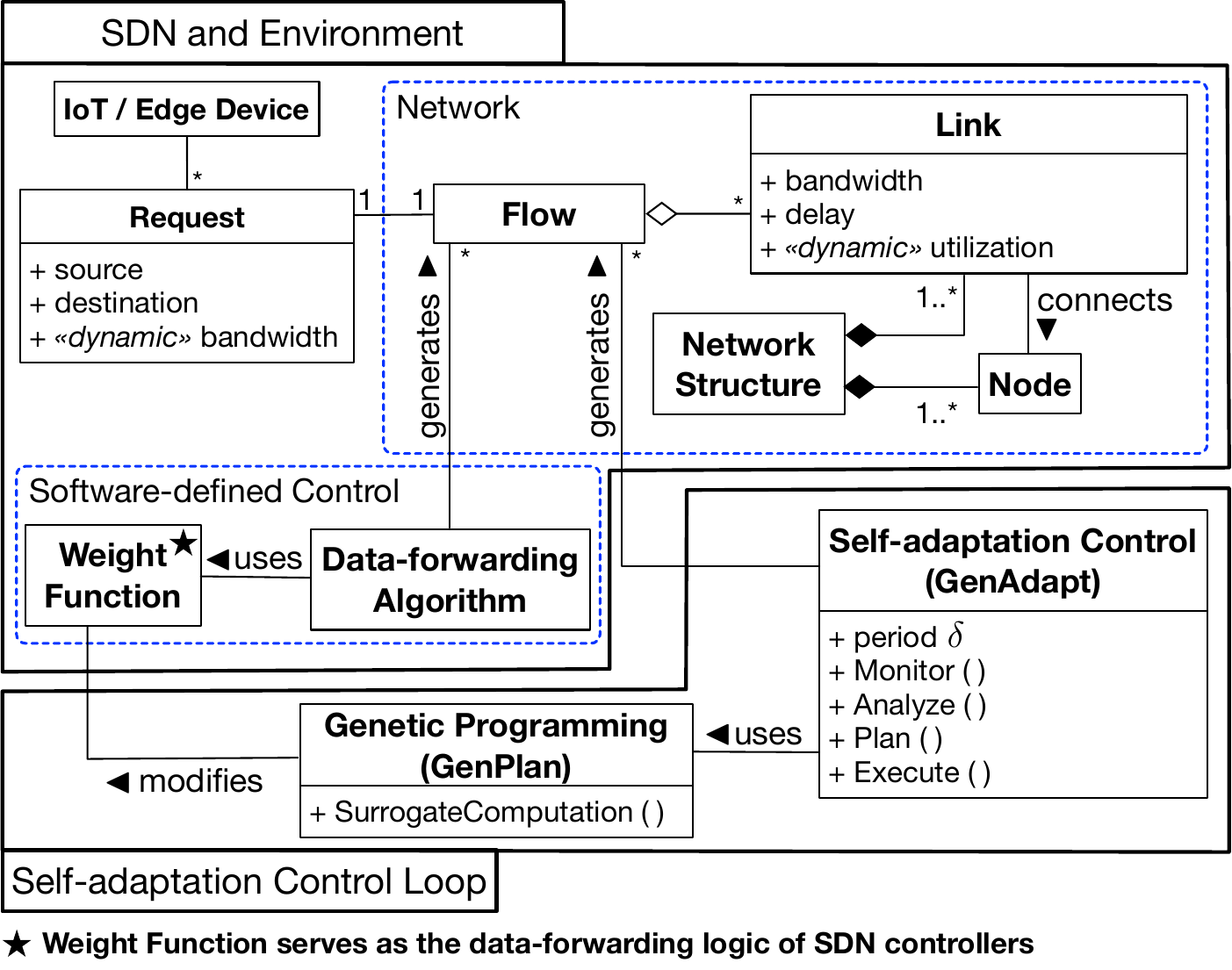}}
	\caption{Domain model for  self-adaptive SDN.}
	\label{fig:model}
\end{figure}

 Figure~\ref{fig:model} shows a domain model consisting of two packages: One package, discussed in Section~\ref{subsec:sysEnv}, captures the main elements of an SDN  and its environment, and the other, discussed in Section~\ref{subsec:selfloop}, specifies the structure of GenAdapt.  As indicated in the figure, the data-forwarding logic of SDN controllers is captured using 
link-weight functions.

\subsection{Domain Model for  Self-adaptive SDN}
\label{subsec:sysEnv}
The \textsf{SDN and Environment} package captures the static structure of a network as well as its dynamic behaviour over time. The environment of an SDN is comprised of IoT and edge devices which are connected to the network and which generate data-transmission requests (requests, for short) that the network needs to fulfil. 

\begin{dfn}[Data-transmission Request]\label{def:datareq}
A data-transmission request $r$ specifies a data stream sent by a network node  $s$  to a network node  $d$. We denote the source node of $r$ by  $r.s$ and the destination node of $r$ by $r.d$. Let $[0..T]$ be a time interval. We denote the (data) bandwidth of $r$ at time $t \in [0..T]$ by $r.bd(t)$. The bandwidth is the amount of data transmitted from $r.s$ to $r.d$ over time. We measure bandwidth in megabits per second (Mbps). The request bandwidth may vary over time (marked as ``dynamic'' in Figure~\ref{fig:model}).
\end{dfn}

In Figure~\ref{fig:model}, the network part of an SDN  is delineated with a dashed rectangle labelled \textsf{Network} and includes the \textsf{Network Structure}, \textsf{Link}, \textsf{Node} and \textsf{Flow} entities. Specifically, a network is a tuple $G=(V,E)$, where $V$ is a set of nodes, and $E \subseteq V \times V $ is a set of directed links between nodes.  Note that each (undirected) link in Figure~\ref{fig:example} represents two directed links. Network links have a nominal maximum bandwidth and  maximum transmission delay assigned to them  based on their physical features and types. The bandwidth of a link is the maximum capacity of the link for transmitting data per second, and the maximum delay specifies the maximum time it takes for data transmission over a link.

\begin{dfn}[Static Properties of a Link]\label{def-statLn}
Let $G = (V, E)$ be a network structure. Each link $e \in E$ has a bandwidth $bw(e)$ and a nominal delay $dl(e)$. We denote by $\mathit{StaticProp}(e)$ the tuple $(bw(e), dl(e))$, indicating the static properties of $e$.  
\end{dfn}

As discussed in Section~\ref{sec:motivate}, a network  handles requests by identifying  a directed path (or a \emph{flow}) in the network to transmit them. Upon the arrival of each request $r$, a network flow (path) $f$ is established to transmit the data stream of $r$ from the requested source $r.s$ to the requested destination $r.d$. Each flow $f$ is a directed path of links that connects $r.s$ to  $r.d$. As shown in Figure~\ref{fig:model},  one flow is created per request. The bandwidth of each flow is equal to that of its corresponding request. Since requests have dynamic bandwidths  (Definition~\ref{def:datareq}), flows have dynamic bandwidths too. We define the throughput of link $e$ at time $t$, denoted by $\mathit{throughput}(e, t)$, as the total of the bandwidths of the flows going through $e$ at time $t$. 

\begin{dfn}[Dynamic Utilization of a Link]\label{def-linkdyn}
Let $G = (V, E)$ be a network, and let $[0..T]$ be a  time interval during which the network is being monitored. At each time instance 
$t \in [0..T]$, each network link $e \in E$ has a utilization $\mathit{util}(e, t)$ computed as follows:
$\mathit{util}(e, t) = \mathit{throughput}(e, t)/\mathit{bw}(e)$.
\end{dfn}

The software-defined control entities  in Figure~\ref{fig:model} include \textsf{Weight Function} and \textsf{Data-forwarding Algorithm}. Data-forwarding algorithms are event-driven and handle requests upon arrival. As discussed in Section~\ref{sec:motivate}, data forwarding typically generates flows based on shortest weighted paths between the source and the destination of a request. Thanks to the SDN architecture, link weights are programmable and can be computed by a weight function that accounts for both the dynamic and the static properties of networks. For example, the GP-derived link-weight function in Figure~\ref{fig:example} (c) uses the dynamic link utilization  $\mathit{util}(e, t)$ and the static parameter $\mathit{threshold}$ to compute link weights. 

\subsection{Generative Self-adaptation}
\label{subsec:selfloop}
In this section, we describe \textit{GenAdapt}, our   self-adaptation control loop, and \textit{GenPlan}, the genetic algorithm used in the planning step of GenAdapt. As discussed earlier,  self-adaptation control has four main steps; these are specified as methods in the self-adaptation control entity in Figure~\ref{fig:model}. GenAdapt, i.e., our self-adaptation loop, runs in parallel with the data-forwarding algorithm. In contrast to the  data-forwarding algorithm which is event-driven, GenAdapt is executed periodically with a period $\delta$,  indicated as an attribute of GenAdapt in Figure~\ref{fig:model}.  The self-adaptation loop periodically monitors the network for congestion as the environment changes, e.g., due to the arrival of new requests.  There is a trade-off between the execution time of GenAdapt and the period $\delta$. In particular, $\delta$ should be small enough so that GenAdapt is executed frequently to detect and handle congestion promptly. At the same time, $\delta$ should be large enough so that frequent executions of GenAdapt do not interfere with other SDN algorithms and applications~\cite{ShinNSB0Z20}. 

\begin{algorithm}[t]
\begin{lstlisting}[style=Alg, frame=0]
Input G : Network structure
Input $\cup_{e \in E}$StaticProp($e$): Static properties of links
Input BestSol: Best Solutions from the previous round
Output F: Optimized flows
Output W: Optimized weight function

for every time step $i$ $\in \{1, \ldots, n\}$ do 
?\vrule? F$_i$, $\cup_{e \in E}$util($e$, $i\cdot \delta$) $\leftarrow$ Monitor() //Dynamic data from SDN Sim
?\vrule? maxUtil = $\mathit{Max}\{\mbox{util}(e, i\cdot \delta)\}_{e \in E}$ 
?\vrule? if maxUtil $>$  $\mathit{threshold}$  then
?\vrule? ?\vrule? F, W $\leftarrow$ GenPlan(G,  $\cup_{e \in E}$StaticProp($e$), F$_i$, BestSol)
?\vrule? ?\vrule? Apply F $\mbox{and}$ W to the SDN data-forwarding algorithm
?\vrule? $\mathbf{end}$
$\mathbf{end}$
\end{lstlisting}
\caption{Self-adaptation loop to resolve congestion by learning a new weight function for SDN data-forwarding ---GenAdapt.}
\label{alg:self-adapt}
\end{algorithm}

GenAdapt, shown in Algorithm~\ref{alg:self-adapt}, executes at every time step $i{\cdot} \delta$ ($i=0, 1, \ldots$).  The \emph{monitor} step (line 8) fetches the set $F_i$ of flows at time $i \cdot \delta$ and the utilization $\mathit{util}(e, i\cdot \delta)$ of every  link $e$ at $i \cdot \delta$. The \emph{analyze} step (lines 9-10) determines  whether the network is congested, i.e., whether adaptation is needed. A network is congested if there is a link $e \in E$ which is utilized above a certain threshold~\cite{Lin:16,Akyildiz:14}. To detect congestion, the maximum utilization of all the links is compared with $\mathit{threshold}$, which is a fixed parameter of the SDN. 

If the network is congested, GenAdapt calls \textit{GenPlan} (line 11). GenPlan, shown in Algorithm~\ref{fig:gp}, is our GP algorithm which we discuss momentarily. The output of GenPlan is a new  link-weight function as well as a modified set of flows where a minimal number of flows have been re-routed.  The new flows and the optimized link-weight function are then applied to the SDN under analysis (line 12).  Note that the only  change that needs to be done to the network is re-routing a typically small number of flows to eliminate congestion.

 GenPlan (Algorithm~\ref{fig:gp}) generates link-weight functions that optimize a fitness function characterizing the desired network-flow properties. The link-weight functions are specified  in terms of static link properties ($\mathit{bw}(e)$ and $\mathit{dl}(e)$), dynamic link utilization ($\mathit{util}(e,t)$) and utilization threshold. Note that GenPlan always reuses in its initial population half of the best solutions (candidate weight functions) generated by its previous invocation and stored in a variable named BestSol. Reusing as bootstrap some of the best solutions from the previous round ensures continuity and incrementality in learning the weight functions across multiple rounds of self-adaptation.

Before we describe the GenPlan algorithm in detail, we define three sets, OldFlows, BadFlows, and NewFlows, that are utilized by GenPlan. 
Specifically, \emph{OldFlows} is the set of all flows when a congestion is detected; \emph{BadFlows} is a subset of OldFlows that should be re-routed to resolve network congestion; and,   \emph{NewFlows} is the set of the new flows computed after re-routing those from BadFlows as well as the original flows that did not cause congestion (i.e., those in $\text{OldFlows}\setminus\text{BadFlows}$). OldFlows is an input to GenPlan, and BadFlows and NewFlows are computed by GenPlan in order to resolve congestion, as we describe below.

 GenPlan  starts by selecting a number of flows, BadFlows, from a congested set of flows, OldFlows, using the FindFlowsCausingCongestion algorithm shown in Algorithm~\ref{fig:removeflows}. Following the standard steps of GP,  GenPlan creates an initial population $P_0$ (line 13) containing a set of possible weight functions (individuals). For every individual $\omega \in P_0$, we call ComputeSurrogate, shown in Algorithm~\ref{fig:surrogate}, to re-route the flows in BadFlows when $\omega$ is used to compute weights of the network links (line 19 of GenPlan). Specifically, for each  $\omega$,   ComputeSurrogate  generates the set NewFlows which includes (1)~the flows corresponding to those in BadFlows, but re-routed based on $\omega$; and (2)~the original flows that did not cause congestion (i.e., $\text{OldFlows}\setminus\text{BadFlows}$).  The set NewFlows is needed to compute the fitness value for $\omega$ (line 20).  The fitness function aims to determine how close an individual is to resolving congestion. Our fitness function combines three criteria as we elaborate momentarily. GenPlan evolves the population by breeding and generating a new offspring population (line 16). The breeding and evaluation steps are repeated until a stop condition is satisfied. Then, GenPlan  returns the link-weight function with the lowest fitness (BestW) and its corresponding flows (line 27).

The ComputeSurrogate algorithm (Algorithm~\ref{fig:surrogate}), which is called by GenPlan, estimates the flows for each candidate weight function $\omega$, assuming that the flow bandwidths obtained at the monitoring step are constant over the duration of $\delta$. ComputeSurrogate first updates the utilization and weight values for each link, assuming that BadFlows are absent (lines 8-11). The new weights are then used to re-route each flow $f$ in BadFlows by identifying a flow $f'$ as the new shortest path (line 13). After creating  each new flow $f'$, the utilization and weight values for each link on  $f'$ are updated (lines 14-17). Note that, because of the assumption that flow bandwidths remain fixed during $\delta$, the utilization-per-link $util(e)$ in ComputeSurrogate is not indexed by time.

\begin{algorithm}[t]
\begin{lstlisting}[style=Alg,frame=0]
Input G: Network structure
Input $\cup_{e \in E}$StaticProp($e$): Static properties of links
Input OldFlows: Current network flows
Input BestSol: Best Solutions from the previous round
Output W: New weight function 
Output NewFlows: New network flows 

t = 0;
BadFlows = FindFlowsCausingCongestion(OldFlows, G);
Flows = OldFlows $\setminus$ BadFlows
while not(stop_condition) do 
?\vrule? if (t == 0)
?\vrule? ?\vrule? P$_0$ = InitialPopOfWeightFormulas() $\cup$ BestSol
?\vrule? ?\vrule? OffSprings = P$_0$
?\vrule? else
?\vrule? ?\vrule? OffSprings = Breed(P$_t$)
?\vrule? $\mathbf{end}$
?\vrule? for $\omega \in$ OffSprings do
?\vrule? ?\vrule? NewFlows = ComputeSurrogate(Flows, BadFlows, G, $\omega$) 
?\vrule? ?\vrule? $\omega$.Fit = Evaluate (G, NewFlows, OldFlows)
?\vrule? $\mathbf{end}$
?\vrule? P$_{t+1}$ = OffSprings
?\vrule? t = t + 1
$\mathbf{end}$
BestW = BestSolution(P$_0$, $\ldots$, P$_t$)
bestFlows = ComputeSurrogate(Flows, BadFlows, G, BestW) 
return BestW, bestFlows
\end{lstlisting}
\caption{Generating a new  link-weight function and a congestion-free set of flows using Genetic Programming ---GenPlan.}
\label{fig:gp}
\end{algorithm}

Following standard practice for expressing meta-heuristic search problems~\cite{Harman2010SearchBS}, we define the representation, the fitness function, and the genetic operators underlying GenPlan.

\textbf{Representation of the Individuals.} An individual represents some weight function induced by the following grammar rule:\\[0em]

        \fbox{\parbox{\dimexpr\linewidth-9\fboxsep-2\fboxrule\relax}{\centering exp ::= 
        exp\,$+$\,exp\,$|$
        exp\,$-$\,exp\,$|$
        exp\,$*$\,exp\,$|$
        exp\,$/$\,exp\,$|$
        const\,$|$
        StaticProp\,$|$ 
       DynamicVar\,$|$ 
        param}}
\\

In the above, the symbol $\mid$ separates alternatives, \texttt{const} is an ephemeral random constant generator~\cite{Veenhuis13},  \texttt{StaticProp} are static link properties (Definition~\ref{def-statLn}), \texttt{DynamicVar} is the link utilization  (Definition~\ref{def-linkdyn}), and \texttt{param} is some network parameter.  
The formula in Figure~\ref{fig:example}(c) can be generated by this grammar rule and is thus an example individual in GenPlan. Note that the SDN data-forwarding algorithm assumes that link values are integers.

\begin{figure}[t]
\makebox[\textwidth][c]{\includegraphics[width=0.3\textwidth]{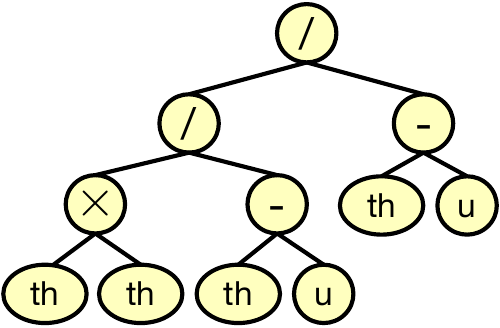}}
	\caption{A parse tree for an example link-weight formula.}
	\label{fig:parse-tree}
\end{figure}

Each individual is constructed and 
manipulated as a parse tree. For example, Figure ~\ref{fig:parse-tree} shows the parse tree corresponding to an example link-weight formula, $\frac{th^2}{ (th - u)^2}$,  where \textit{u} is a shorthand for $util(e, t)$ and \textit{th} is a shorthand for threshold. The initial population of GenPlan is generated by randomly building parse trees using the grow method~\cite{poli2008field} (i.e., the root and inner nodes are labelled by the mathematical operations, and the leaves are labelled by variables, constants or parameters specified in the grammar). As discussed earlier when GenPlan is called for the first time,  the initial population is generated randomly. For subsequent calls to GenPlan, half of the initial population is generated and the other half is reused from the best elements in the last population generated by the previous invocation of GenPlan.

\begin{algorithm}[t]
\begin{lstlisting}[style=Alg, frame=0]
Input G: Network structure
Input Flows: Current flows
Output BadFlows: A subset of Flows causing congestion 

BadFlows = $\emptyset$
while G is congested do 
?\vrule? Let $e$ $\mbox{be}$ the most congested link
?\vrule? Let $f \in$ Flows such that  $e \in f$ // selected randomly
?\vrule? BadFlows = BadFlows$\cup \{f\}$
?\vrule? Flows = Flows$\setminus \{f\}$
$\mathbf{end}$
return BadFlows
\end{lstlisting}
\vspace*{.25cm}
\caption{Selecting a subset of flows whose removal will resolve congestion ---FindFlowsCausingCongestion.}
\label{fig:removeflows}
\vspace*{-.1cm}
\end{algorithm}

\begin{algorithm}[t]
\begin{lstlisting}[style=Alg, frame=0]
Input G: Network structure
Input Flows: Current congestion-free set of flows
Input BadFlows: Flows that have to $\mbox{be}$ re-routed 
Input $\omega$: Candidate weight function
Output NewFlows: BadFlows re-routed using $\omega$

NewFlows $= \emptyset$
for $e \in E$ do
?\vrule? util($e$) $\leftarrow$ sum of the bandwidths of $f \in$ Flows s.t. $e \in f$
?\vrule? Compute weight $\mbox{for}$ $e$ using $\omega$
$\mathbf{end}$
for $f \in$ BadFlows do
?\vrule? $f'$ $\leftarrow$ shortest weighted path from  source to destination of $f$ 
?\vrule? for $e \in f'$ do
?\vrule? ?\vrule? util($e$) = util($e$) + bandwidth of $f'$ 
?\vrule? ?\vrule? update the weight $\mbox{of}$ $e$ using $\omega$
?\vrule? $\mathbf{end}$
?\vrule? NewFlows = NewFlows$\cup \{f'\}$
$\mathbf{end}$
return NewFlows $\cup$ Flows
\end{lstlisting}
\caption{Re-routing congested flows (BadFlows) by computing shortest paths based on a candidate weight function ($\omega$) ---ComputeSurrogate.}
\label{fig:surrogate}
\end{algorithm}

\textbf{Fitness Function.} We propose a fitness function for resolving network congestion in SDNs.
Our fitness function is hybrid and  combines the following three metrics: (1)~Maximum link utilization across all the network links ($\mathit{Fit1}$).
For each individual weight function $\omega$, GenPlan computes the set NewFlows of flows based on link weights generated by $\omega$.  The $\mathit{Fit1}$ metric computes the utilization value of the most utilized network link, considering the flows in NewFlows. If $\mathit{Fit1}$ is higher than the threshold, then the network is congested. Hence, we are interested in individuals whose $\mathit{Fit1}$ is less than the threshold. (2)~The cost of re-routing network flows measured as the number of  link updates, i.e., insertions and deletions, required to reconfigure the network flows ($\mathit{Fit2}$). In GenPlan, OldFlows is the current set of congested flows, and as mentioned above, NewFlows is the set of flows computed for each individual $\omega$. We compute $\mathit{Fit2}$ as the edit distance between each pair of flows $f \in$ OldFlows  and $f'\in$ NewFlows  such that $f$ and $f'$ are both related to the same request.  Specifically, the distance between two flows $f$ and $f'$ is measured as the longest common  subsequence (LCS) distance of two paths~\cite{Cormen:09} by counting the number of link insertions and link deletions required to transform $f$ into $f'$. We note that this metric has previously been used as a proxy for the reconfiguration cost of network flows~\cite{ShinNSB0Z20}. (3)~The total data transmission delay generated by the new flows ($\mathit{Fit3}$). The $\mathit{Fit3}$ metric is computed as the sum of all the delay values, i.e., $\mathit{dl(e)}$, of the links that are utilized by the new flows (i.e., the NewFlow set).  The larger this value, the higher the overall transmission delay induced by NewFlows. The last metric allows us to penalize candidates that generate longer flows compared to those that generate shorter ones. 

The $\mathit{Fit1}$, $\mathit{Fit2}$ and $\mathit{Fit3}$ metrics have different units of measure and ranges. Thus, before combining them, we normalize them using the well-known rational function  $\bar{x} = x/(x + 1)$\cite{Arcuri:10}. This function provides good guidance to the search for minimization problems compared to other alternatives.

Among the three metrics, lowering  $\mathit{Fit1}$ below the congestion threshold takes priority; if a candidate solution is unable to resolve congestion, then we are not interested in the other two metrics. Once $\mathit{Fit1}$ is below the threshold, we do not want to lower $\mathit{Fit1}$ any further, since we want the network optimally utilized but not congested. Instead, we are interested in lowering the cost and delay metrics.  We denote the normalized forms of our three metrics by $\overline{\mathit{Fit1}}$, $\overline{\mathit{Fit2}}$ and $\overline{\mathit{Fit3}}$, respectively, and define the following overall fitness function to combine the three: 

\vspace*{.4em}

$\begin{array}{l}
Fit =
\begin{cases}
\overline{\mathit{Fit1}} +  2 &  \text{(1) If $\mathit{Fit1} \geq \mathit{threshold}$}\\
\overline{\mathit{Fit2}} + \mathit{\overline{Fit3}} &  \text{(2) If  $\mathit{Fit1} < \mathit{threshold}$} \\
\end{cases} 
\end{array}$
\vspace*{.4em}

Given a candidate weight function $\omega$,  evaluating $\mathit{Fit}(\omega)$ always yields a value in $[0..3]$:  $\mathit{Fit}(\omega) \geq 2$ indicates that $\omega$ is not able to resolve congestion; and $\mathit{Fit}(\omega) <2 $ indicates that $\omega$ can resolve congestion, and its fitness determines how well  $\omega$ is doing in reducing  cost and delay. Note that in our fitness function defined above, cost and delay are equally important; we do not prioritize either one. If desired, one can modify the above function by adding coefficients to $\overline{\mathit{Fit2}}$  and $\overline{\mathit{Fit3}}$ to prioritize cost over delay or vice versa.

\textbf{Genetic Operators.}  
We use one-point crossover~\cite{onepointcrossover}. It randomly  selects  two  parent individuals. It then randomly selects one sub-tree in each parent, and  swaps the  selected  sub-trees  resulting  in  two  children. For the mutation operator, we  use  one-point  mutation~\cite{poli1998schema}  that mutates a child individual by randomly selecting one sub-tree and replacing it with a randomly generated tree, which is generated using the initialization procedure. For the parent selection operator, we use tournament selection~\cite{Luke:13}.

\textbf{Symbolic Complexity of GenPlan.} The most computationally expensive task when computing our fitness function is finding  the shortest weighted path from the source switch to the destination switch for each candidate individual. The complexity of finding the shortest weighted path is O(($E+V$)$\log V$), where $E$ is the number of links and $V$ is the number of switches in the network~\cite{CLR}.

\section{Tool Support}\label{sec:tool}

\begin{figure*}[tb]
\makebox[\textwidth][c]{\includegraphics[width=0.8\textwidth]{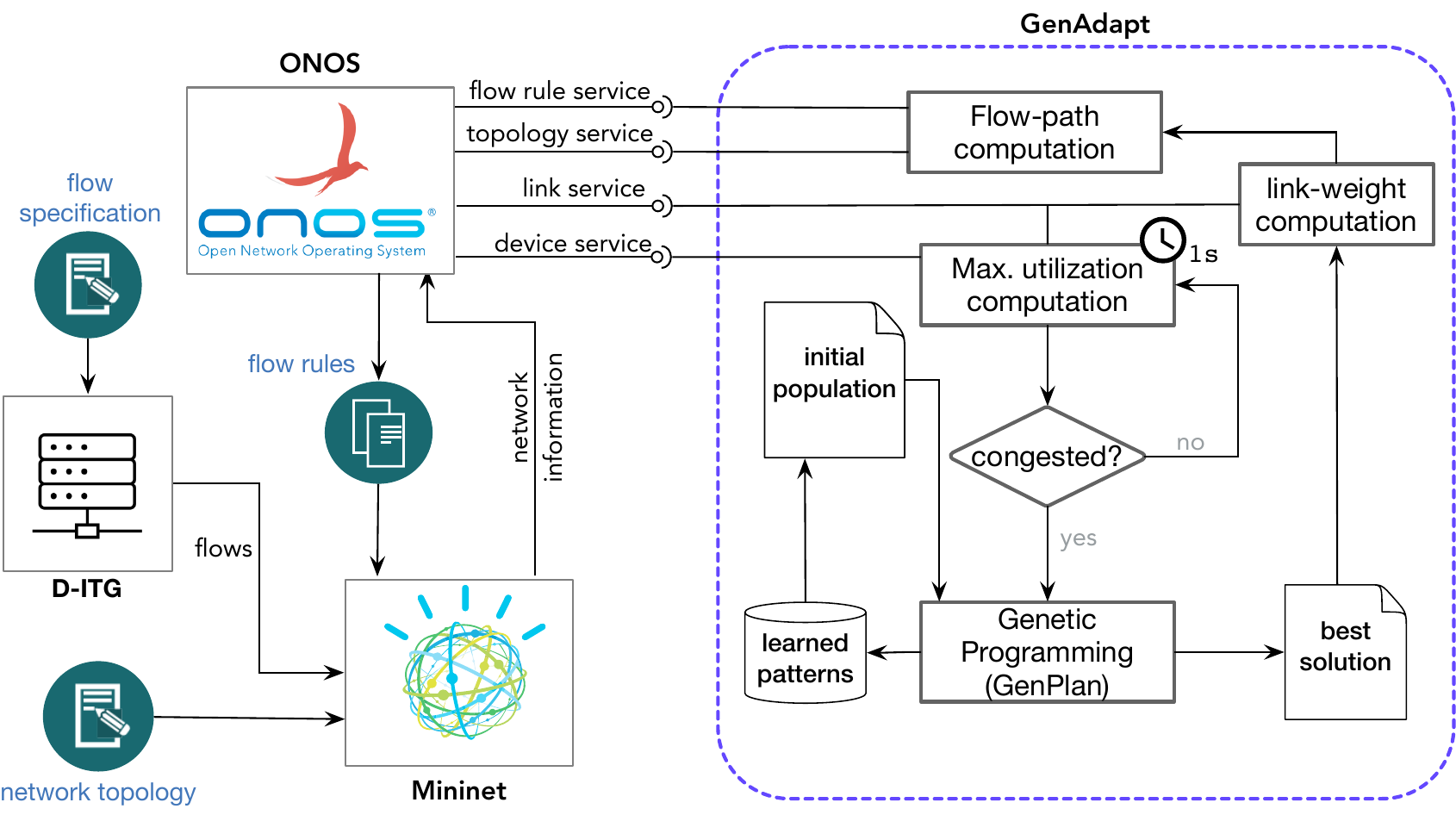}}
	\caption{Tool Architecture.}
	\label{fig:architecture}
\end{figure*}

In this section, we describe the tool that we have developed to build self-adaptivity into SDN data forwarding. Figure~\ref{fig:architecture} shows the architecture of our tool support. The tool is composed of four main modules: A traffic generation component (D-ITG)~\cite{Botta:12}, a network emulator (Mininet)~\cite{Lantz:10}, an SDN controller (ONOS)~\cite{Berde:14} and an implementation of our self-adaptation framework (GenAdapt). Below, we discuss each of these modules.

We use \emph{Distributed Internet Traffic Generator (D-ITG)}~\cite{Botta:12} -- a network traffic generator platform -- to generate flows based on a specification provided by users (e.g., network administrators). These flows are meant for simulating real-world traffic in real networks. D-ITG can generate various flows with different protocols, bandwidth sizes, and temporal features communicated between the nodes in a network. D-ITG keeps track of various metrics related to the generated data flows. These metrics include, among others, packet loss, delay and jitter. 

\emph{Mininet} ~\cite{Lantz:10} is a network emulator which is used for creating a realistic virtual network consisting of virtual hosts, switches and links according to a user-provided network topology. The switch type that we are interested in among the alternative choices available in Mininet is the ``Open virtual Switch (vSwitch)''. This switch type is Mininet's default and runs in OpenFlow mode. This makes it possible to control switches of this type using an OpenFlow controller such as ONOS (discussed in the next paragraph). Flow tables are installed on switches for packet forwarding and can be updated based on the instructions sent by the OpenFlow controller which the switches are connected to.

\emph{Open Network Operating System (ONOS)} ~\cite{Berde:14} is a popular, carrier-grade SDN controllers that can be used with either real or simulated networks. We use ONOS as the OpenFlow controller for the virtual networks created by Mininet. Using the OpenFlow protocol~\cite{McKeown:08}, ONOS communicates with the Open vSwitches in a Mininet network. This enables ONOS (i) to obtain a global view of the network by collecting network information and (ii) to manipulate the flow rules as needed in order to improve network performance. ONOS provides a set of services including flow rule service, topology service, link service and device service. All these services can be accessed and programmed via ONOS's API.

Our adaptation framework, \emph{GenAdapt}, is realized as an add-on for ONOS. As discussed in Section~\ref{sec:overview}, GenAdapt consists of four steps: monitor, analyze, plan and execute.

In the first step, our GenAdapt implementation uses ONOS's device service to collect network metrics from the switches. In the second step, which is triggered with a periodicity of one second, GenAdapt uses ONOS's link service alongside the metrics collected in the previous step to compute the utilization of each network link and obtain the maximum link utilization. As discussed in Section~\ref{sec:selfcontrol}, the network is considered congested if the maximum utilization is greater than a certain threshold. Once a congestion is detected, GenAdapt enters the third step and calls GenPlan (Algorithm ~\ref{fig:gp}) to generate a new link-weight function (best solution) as well as a set of new link weights calculated by the best solution. In the fourth and final step, using the link weights calculated in the third step, ONOS's flow rule service computes new shortest weighted paths and updates the flow rules for the switches according to these paths. This step resolves congestion by rerouting some of the flows (specifically, BadFlows in algorithm shown in Algorithm~\ref{fig:removeflows}) based on the new flow rules.

The best solution derived by an invocation of GenPlan (step three) is retained and applied for as long as it has not been replaced by a newer solution (by GenPlan) in response to a future congestion occurrence. The candidate functions created in the last iteration of GP in GenPlan are sorted by their fitness values. The top half of these candidate functions are kept as patterns learned, to be reused by the next invocation of GenPlan. Upon its next invocation, GenPlan creates its initial population as follows: 50\% of the initial population are the patterns learned from the previous invocation and the other 50\% are randomly generated candidate functions. As we argue in Section~\ref{sec:overview}, reusing the patterns learned helps ensure continuity across multiple rounds of self-adaptation.

We implemented GenAdapt based on the open-source version of DICES ~\cite{ShinNSB0Z20}, a self-adaptive packet forwarding application, which itself builds on ONOS's default reactive forwarding application (called \emph{org.onosproject.fwd}). For the GenPlan component, we use the genetic programming module provided by ECJ (version 27) ~\cite{DBLP:conf/gecco/ScottL19}. GenAdapt is available online ~\cite{appedix}.
\vspace*{.3cm}
\section{Empirical Evaluation}
\label{sec:eval}
In this section, we investigate the following research questions (RQs) using open-source synthetic and industrial networks. To answer RQ1, we use the industrial network and eighteen synthetic networks. To answer RQ2, we use six synthetic networks from RQ1. To answer RQ3, we use eight more synthetic networks alongside two of the synthetic networks from RQ1. In total, our evaluation involves one industrial network and 26 synthetic ones.  

\textbf{RQ1 (Effectiveness): } \emph{How effective is GenAdapt in modifying the logic of the SDN data-forwarding algorithm to avoid future congestions?} 
The main novelty of GenAdapt is in attempting to adapt the logic of SDN data forwarding (i.e., the link-weight functions) instead of adapting its output (i.e., the individual flow paths). Through RQ1, we compare GenAdapt with two baseline techniques: (i)~an approach, named DICES~\cite{ShinNSB0Z20}, which, similar to GenAdapt, uses MAPE-K self-adaptation, but optimizes individual flow paths; and (ii)~OSPF  configured using a standard heuristic for setting optimized link weights~\cite{Coltun:08,Cisco:05}. By comparing GenAdapt with these baselines, we investigate whether GenAdapt, when called to resolve an existing congestion, is able to reduce the number of occurrences of congestion in the future, without incurring additional overhead.

\textbf{RQ2 (Transferability):
}
\emph{Can we improve GenAdapt's performance for a larger network by bootstrapping GenAdapt with  best solutions computed on a smaller network that has the same topology as the larger one?} The GenAdapt algorithm (Algorithm~\ref{fig:surrogate}) requires computing shortest weighted paths between different nodes in a network. Hence, the algorithm's execution time increases with network size, i.e., the number of links and nodes. Yet, link-weight functions do not contain information that is specific to the underlying network size. 
We therefore hypothesize that, for networks with similar topology but different sizes, the link-weight functions are likely to be similar, thus making it possible to enhance the performance of GenAdapt by transferring  to larger networks the best control logic learned on smaller networks. To validate this hypothesis, we consider a set of networks with the same topology but different sizes. Before  applying GenAdapt on the larger networks in this set, we initialize (part of) its population using the  best control logic that GenAdapt has learned on the smaller networks in this set.  We then compare the results with those obtained  when GenAdapt's first population is initialized randomly.

\textbf{RQ3 (Scalability):} \emph{Can GenAdapt resolve congestion efficiently as the size of the network and the number of requests increase?} To assess scalability, we evaluate the execution time of GenAdapt as the size of the network and the number of  requests increase.

\textbf{Experimental Setup.}
All experiments were performed on a machine with a 2.5 GHz Intel~Core~i9 CPU and 64 GB of memory. All our experimental material is available online~\cite{appedix}.

\subsection{RQ1 -- Effectiveness}
\label{sec:rq1}
Before answering RQ1, we present the baselines, the study subjects,  the configuration of GenAadapt and the setup of our experiments.

\textbf{Baselines.}  Our first baseline,  DICES, is from the SEAMS literature. Similar to GenAadapt, DICES is self-adaptive, but unlike GenAadapt, it uses a genetic algorithm to modify the individual flows generated by the SDN data-forwarding algorithm.  The second baseline is  OSPF~\cite{Coltun:08,Cisco:05} configured by setting the link weights to be inversely proportional to the bandwidths of the links as suggested by Cisco~\cite{Cisco:05}. The OSPF heuristic link weights are meant to induce optimal flows that eliminate or minimize the likelihood of congestion. OSPF is widely used in real-world systems~\cite{FortzT02} and as a baseline in the literature~\cite{Poularakis:19,Amin:18,Bianco:17,Rego:17,Caria:15,Bianco:15,Agarwal:13}.

\textbf{Study Subjects.} For RQ1, we use: (1)~eighteen synthetic networks, and (2)~an industrial SDN-based IoT network published in earlier work~\cite{ShinNSB0Z20}. For the synthetic networks, we consider two  network topologies: (i)~complete graphs, and (ii)~multiple non-overlapping paths between node pairs. The network in Figure~\ref{fig:example} is an example of the latter topology, where $s_1$ and $s_2$ are connected by three non-overlapping paths.
Since our approach works by changing flow paths, we naturally focus our evaluation on topologies with multiple paths from a source to a destination, thus excluding topologies where adaptation through re-routing is not possible. The topologies that we experiment with, i.e., (i) and (ii) above, are the two extreme ends of the spectrum in terms of path overlaps between node pairs: In the first case, we have complete graphs, where there are as many  overlapping paths as can be between node pairs; and, in the second case, we have no overlapping paths at all. We consider seven complete graphs -- referred to as FULL hereafter -- with five, six, seven, ten, fifteen, twenty two and thirty two nodes, and consider four multiple-non-overlapping path graphs -- referred to as MNP hereafter -- with five, eight, twelve and seventeen nodes. More precisely, one MNP graph has five nodes connecting the designated source and destination with three non-overlapping paths (Figure~\ref{fig:example}); others have eight, twelve and seventeen nodes doing the same with four, five, and six (non-overlapping) paths. Following the suggested parameter values in the existing literature for such experiments~\cite{ShinNSB0Z20}, we set the static properties of the links, namely bandwidth and delay, to 100Mbps and 25ms, respectively.

To instigate changes in the SDN environment, we generate data requests over time and not at once. We space the requests $10$s apart to ensure that all the requests are properly generated and that the network has some time to stabilize, i.e.,  we send requests at $0$s, $10$s, $20$s, $30$s, and so on. For each FULL network, we fix a source and a destination node, and  generate either three or four requests every $10$s. For each MNP network, we generate every $10$s two requests between the end-nodes connected by multiple paths. The reason why we generate fewer requests in the MNP networks is because there are fewer paths compared to the FULL networks. For a given network (FULL or MNP), the generated requests have the same bandwidth. The request bandwidth for each network is selected such that  some congestion is created starting from $10$s. The request bandwidths are provided online~\cite{appedix}. We denote our synthetic networks by \hbox{FULL($x$, $y$)} and MNP($x$, $y$), where $x$ is the number of nodes and $y$ is the number of requests generated every $10$s. The eighteen synthetic networks that we use in RQ1 are fourteen FULL networks, once with three requests and once with four requests generated every $10$s, and four MNP networks with two requests generated every $10$s.

Our industrial subject is an emergency management system (EMS)  from the literature~\cite{ShinNSB0Z20}. EMS represents a real-world application of SDN in a complex IoT system. This subject contains seven switches and 30 links with different values for the links' static properties. The EMS subject includes a traffic profile characterizing anticipated traffic at the time of a natural disaster (e.g., flood), leading to congestion in the network of the monitoring system. In particular, the network is used for transmitting 28 requests capturing different data stream types such as audio, video and sensor and map data. The static properties of the network links and the data-request sizes are available online~\cite{appedix}.\footnote{Due to platform differences, we could not reproduce the congestion reported in~\cite{ShinNSB0Z20}. So, we increased the request bandwidths by 30\% to reproduce congestion.} 

\textbf{Experiments and metrics.} RQ1 has two goals: (G1)~determine whether, by properly modifying the link-weight function, GenAdapt is able to reduce the number of times congestion happens, and (G2)~determine whether GenAdapt can respond quickly enough so that prolonged periods of congestion can be avoided.

The synthetic subjects are best used for achieving G1, since the requests in these subjects are generated with time gaps as opposed to all at once, which is the case in EMS (industry subject). This characteristic of the synthetic subjects creates the potential for congestion to occur multiple times, in turn allowing us to achieve G1. For G1, we compare GenAdapt with DICES; comparison with the OSPF baseline does not apply, since OSPF's heuristic cannot avoid congestion for our synthetic subjects, and neither can it resolve congestion. To compare GenAdapt with DICES, we keep track of how many times congestion occurs during our simulation, the execution time of each technique to resolve each congestion occurrence, and the total time during which the network is in a congested state. In addition, we measure packet loss, which is a standard metric for detecting congestion in network systems. Packet loss is measured as the number of dropped packets divided by the total number of packets in transit across the entire network during simulation.  The simulation time for each synthetic subject was set to end $10$s after the generation of the last request. Recall that, in our synthetic subjects,  the requests are sent at intervals of $10$s and as long as the number of paths in the underlying network is sufficient to fulfill the incoming requests.

To achieve G2, we use EMS in order to compare GenAdapt with DICES and OSPF in terms of handling the congestion caused by requests arriving at once. Since DICES and GenAdapt are self-adaptive, they monitor EMS and attempt to resolve congestion when it occurs.  In the case of OSPF, however, the heuristic link weights are meant to reduce the likelihood of congestion and packet loss.  For this comparison, we report the total packet loss recorded by each of the three technique over a fixed simulation interval of $5$min, and also whether, or not,  GenAdapt and DICES were able to resolve the congestion in EMS within the simulation time.

\textbf{Configuring GenAdapt.} Table~\ref{tab:parameters} shows the configuration  parameters used for GenAdapt. For the mutation and crossover rates, the maximum tree depth and the tournament size, we chose  recommendations from either the GP literature~\cite{poli2008field,LukeP06} or ECJ's documentation~\cite{DBLP:conf/gecco/ScottL19}. We set $\delta$ to $1$s since this is the smallest monitoring period permitted by our simulator. We use the utilization threshold given in the literature~\cite{ShinNSB0Z20,Akyildiz:14,Lin:16}.  Since the range for link-utilization values is [$0$\% .. $100$\%], we set the minimum and maximum of the constants in GP individuals to $0$ and $100$, respectively.

\begin{table}
\caption{Parameters of GenAdapt.}
\label{tab:parameters}
\begin{center}
\scalebox{.75}{
\begin{tabular}{p{4cm} p{0.5cm}||p{8cm} p{0.5cm}}
\toprule
Mutation rate  &  0.1 & Utilization threshold  &  $80$\% \\
\hline
Crossover rate &  0.7 & Minimum of the constant in the GP grammar &   0 \\
\hline
Maximum depth of GP tree &  15 & Maximum of the constant in the GP grammar   &  100 \\
\hline
Tournament size &  7  &  Interval between invocations of GenAdapt ($\delta$) &  $1$s   \\
\hline

Population size &  10 & 
\\
\bottomrule
\end{tabular}}
\end{center}
\end{table}

To determine the population size and the number of generations for GP, we note that, ideally, GenAdapt should not take longer than $\delta$ to execute. The execution time of GenAdapt is likely shorter when it is called in the first rounds of request generation than in the later rounds, since there are fewer flows in the early rounds of our synthetic subjects. Hence, using the fixed time limit of $1$s to stop GenAdapt is not optimal. Instead of using a time limit, we performed some preliminary experiments on our synthetic subjects to configure population size and the number of generations for the two topologies of FULL and MNP. For both topologies,  we opt to use a small population size (i.e., $10$). As for the number of generations, for the FULL networks, we stop GenAdapt when the fitness function falls below two (i.e., when congestion is resolved but the solution is not necessarily optimized for delay and cost) or when 200 generations is reached. This will ensure that GenAdapt's execution time does not exceed $1$s for our FULL networks. The MNP networks are, however, sparser and we are able to increase the number of generations while keeping the execution time below $1$s. Specifically, for MNP networks with five nodes, we use 300 generations; and, for the ones with eight, twelve and seventeen nodes we use 500 generations.  For EMS, the topology is more similar to a full graph, and as such, we use the same configuration for EMS as that for FULL networks.

\textbf{Formulas Generated by GenPlan.}
We charaterize the structure and  shape of the formulas generated by GenPlan using the following metrics: (1)~the number of operators, which represents the number of ``$+$'', ``$-$'', ``$*$'' and ``$/$'' symbols within a formula; (2)~the number of variables, which refers to the number of \textsf{const}, \textsf{StaticProp}, \textsf{DynamicVar}, and \textsf{param} elements within a formula; and, (3)~ the depth of the parse tree corresponding to a formula. As each formula is represented by a parse tree, the number of operators corresponds to the number of non-leaf nodes, and the number of variables corresponds to the number of leaf nodes in the tree. For example, the link-weight formula $\frac{th^2}{ (th - u)^2}$ , with its corresponding parse tree illustrated in  Figure~\ref{fig:parse-tree}, contains 5 operators, 6 variables and has a parse-tree depth of 4. 
 We randomly selected 20 best individuals  resulting from 20 runs of GenPlan.  In these individuals, the number of operators ranges from 0 to 28, with an average of 6.6 and a median of 4. The number of variables, which also represents the number of leaves in the parse tree, ranges from 1 to 29, with an average of 7.6 and a median of 5. The depth of the parse trees varies from 1 to 15, with an average depth of 4.3 and a median of 4. In GenPlan, individuals that  result in a divide by zero are dropped, since they are invalid.

\begin{figure}[t]
	\centerline{\includegraphics[width=.8\columnwidth]{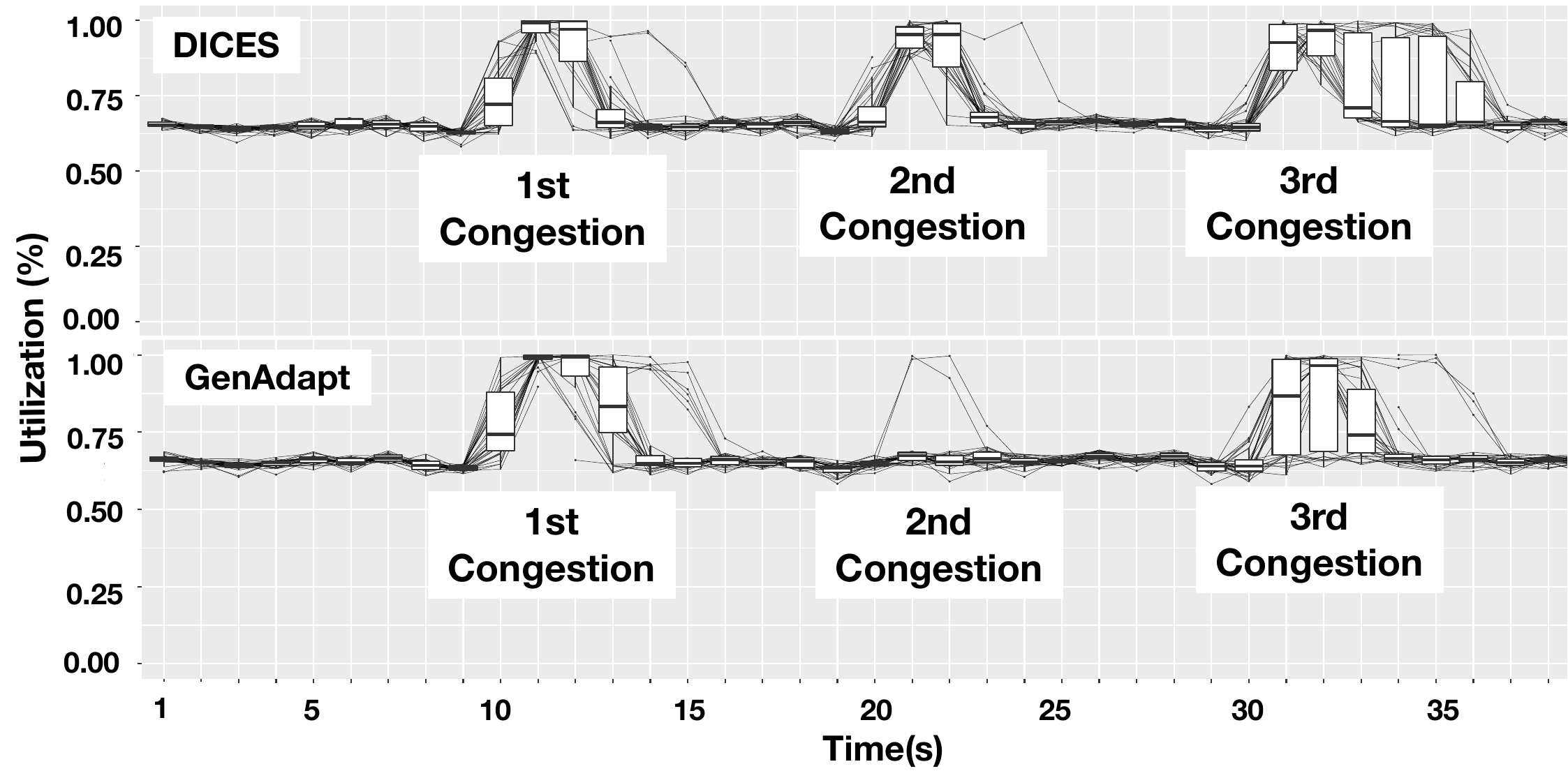}}
	\caption{Comparing network utilization values over time obtained from 30 runs of DICES and GenAdapt for resolving congestion in an example synthetic subject: MNP(8, 2).}
	\label{fig:rq1}
\end{figure}

\textbf{Results.} 
As an example, Figure~\ref{fig:rq1} shows the network utilization over time when GenAdapt and DICES are used to resolve congestion for the synthetic network MNP(8, 2). Two simultaneous network requests are sent at 0s, 10s, 20s and 30s, leading to congestion at 10s, 20s, and 30s.  The network is congested when the utilization value is above 0.8 (i.e., the utilization threshold). Out of 30 runs for DICES, 18 runs record three congestion occurrences and 12 runs record four. In contrast, out of 30 runs for GenAdapt, six runs record only one congestion occurrence, 16 runs record two, seven runs record three, and only one run records four congestion occurrences.  Note that three congestion occurrences at 10s, 20s, and 30s are visible in the figure. However, in the simulated environment, similar to the physical world, there are some small time gaps between the arrivals of the two requests generated each time, and in addition, there are small fluctuations in flow bandwidths over time. Hence, the monitoring step of GenAdapt or DICES may detect congestion twice (i.e., once per request arrival), instead of only once and after the arrival of both requests. In total, for the example in Figure~\ref{fig:rq1}, the average number of congestion occurrences is 2.1 with GenAdapt and 3.4 with DICES. In addition, with DICES, the average of the total time that the network remains congested  is  7.6s, while with GenAdapt, this is reduced to 5.57s.

\begin{table}[th]
	\caption{Statistical-test results comparing the average number of congestion occurrences, the duration of congestion, the execution-time and the packet-loss values obtained by 30 runs of GenAdapt versus DICES for our eighteen synthetic subjects.}
	\label{tab:GpvsDices}
	\centering
	\scalebox{0.6}{
	\begin{tabular}{||c|c|c|c||c|c|c||c|c|c||}
	\toprule
	  & \multicolumn{3}{c||}{ \textbf{\emph{FULL(5,3)}}} & 
	  \multicolumn{3}{c||}{ \textbf{\emph{FULL(5,4)}}} &  
	  \multicolumn{3}{c||}{ \textbf{\emph{FULL(6,3)}} -- {\color{red}\textbf{40\% of DICES runs failed}}}  \\
	 \hline
 & \textbf{avg(G-D)$^*$}& \textbf{p-value} & \textbf{$\bf{\hat{A}_{12}}$} &
 \textbf{avg(G-D)}& \textbf{p-value} & \textbf{$\bf{\hat{A}_{12}}$} &
 \textbf{avg(G-D)}& \textbf{p-value} & \textbf{$\bf{\hat{A}_{12}}$}\\
 \hline
\# Congestion & 3.1 -- 3.2 & 0.98 & 0.5(N) & 
3.8 -- 4.4 & 0.193 & 0.59(S) & 
3.5 -- 4.5 & 0.003 & 0.74(L) \\

Congestion Duration (s) & 6.2 -- 7.7 & 0.0919  & 0.62(S)& 
9.53 -- 12 & 0.145  & 0.61(S) & 
7.93 -- 12 & 1.69E-06  & 0.86(L) \\
Packet Loss (\%) & 32.1 -- 32.33 & 0.001 & 0.75(L) & 
24.86 -- 24.99 & 0.535 & 0.45(N) & 
24.94 -- 31.73 & 7.96E-11 & 0.94(L) \\ 

Exec Time (ms)  & 140.86 -- 392.96 & <2.2E-16 &0.98(L) & 
271.95 -- 440.87 & <2.2E-16 & 0.87(L) & 
302.21 -- 701.06 & <2.2E-16 & 0.9(L)  \\
\bottomrule
  & 

	  \multicolumn{3}{c||}{ \textbf{\emph{FULL(6,4)}} -- {\color{red}\textbf{
	  33.3\% of DICES runs failed}}} & 
	  \multicolumn{3}{c||}{ \textbf{\emph{FULL(7,3)}} -- {\color{red}\textbf{66.6\% of DICES runs failed}}}  & 
  \multicolumn{3}{c||}{ \textbf{\emph{FULL(7,4)}}}
	  \\
\hline
& \textbf{avg(G-D)}& \textbf{p-value} & \textbf{$\bf{\hat{A}_{12}}$} &
 \textbf{avg(G-D)}& \textbf{p-value} & \textbf{$\bf{\hat{A}_{12}}$} &
 \textbf{avg(G-D)}& \textbf{p-value} & \textbf{$\bf{\hat{A}_{12}}$} \\  
\hline
\# Congestion & 
3.8 -- 5.4 & 5.39E-06 & 0.87(L) &
4.1 -- 6 & 9.70E-08 & 1(L) & 
2.9 -- 3.1 & 0.063 & 0.59(S)  \\

Congestion Duration (s) &
10.03 -- 14.47 & 5.98E-05  & 0.8(L) & 
8.03 -- 16.63 & 1.35E-11 & 1(L) & 
5 -- 5.97 & 0.0157  & 0.68(M) \\ 

Packet Loss (\%)  & 
28.4 -- 31.64 & 1.53E-05 & 0.83(L) & 
16.6 -- 19.7 & 0.145 & 0.61(S) & 
10.13 -- 19.31 & 8.08E-07 & 0.87(L) \\

Exec Time (ms) & 
956.47 -- 802.48 & <2.2E-16 & 0.85(L) & 
295.94 -- 949.2 & <2.2E-16 & 0.97(L) & 
226.40 -- 576.05 & <2.2E-16 & 0.99(L)\\
\bottomrule
\bottomrule
  & 

	  \multicolumn{3}{c||}{ \textbf{\emph{FULL(10,3)}}} &  
	  \multicolumn{3}{c||}{ \textbf{\emph{FULL(10,4)}} -- {\color{red} \textbf{ 10\% of DICES runs failed}}} & 
	  \multicolumn{3}{c||}{ \textbf{\emph{FULL(15,3)}}} 
	  \\
\hline
& \textbf{avg(G-D)}& \textbf{p-value} & \textbf{$\bf{\hat{A}_{12}}$} &
 \textbf{avg(G-D)}& \textbf{p-value} & \textbf{$\bf{\hat{A}_{12}}$} &
 \textbf{avg(G-D)}& \textbf{p-value} & \textbf{$\bf{\hat{A}_{12}}$} \\  
\hline
\# Congestion & 
3.7 -- 4 & 0.464 & 0.58(S) & 
4 -- 5.9 & 9.21E-11 & 0.96(L) & 
1.87 -- 5 & 2.757e-09 & 0.9 (L) \\

Congestion Duration (s) & 
7.5 -- 8.47 & 0.029  & 0.65(S) & 
8.9 -- 16.23 & 9.53E-11  & 0.98(L) & 
2.73 -- 5.77 & 1.017E-06 & 0.86 (L) \\ 

Packet Loss (\%)  &
8.75 -- 13.02 & 0.006 & 0.71(M) & 
30.17 -- 34.41 & 1.61E-10 & 0.98(L) & 
14.16 -- 15.29 & 3.077E-08 & 0.92 (L)  \\

Exec Time (ms) & 
564.88 -- 814.8 & 4.65E-16 & 0.81(L) & 
596.31 -- 1365.28 & <2.2E-16 & 0.86(L) & 
732.92 -- 1574.83  & < 2.2E-16 & 0.89(L) \\
\bottomrule
\bottomrule
  &
 \multicolumn{3}{c||}{ \textbf{\emph{FULL(15,4)}}}    & 
 \multicolumn{3}{c||}{ \textbf{\emph{FULL(22,3)}}} & 
 \multicolumn{3}{c||}{ \textbf{\emph{FULL(22,4)}}}
	  \\
\hline
& \textbf{avg(G-D)}& \textbf{p-value} & \textbf{$\bf{\hat{A}_{12}}$} &
 \textbf{avg(G-D)}& \textbf{p-value} & \textbf{$\bf{\hat{A}_{12}}$} &
 \textbf{avg(G-D)}& \textbf{p-value} & \textbf{$\bf{\hat{A}_{12}}$}  \\  
\hline
\# Congestion & 
2.53 -- 3 & 0.0002 & 0.73 (M) & 
 2.1 --5  & 1.681E-09 & 0.92(L) & 
 2.3 -- 3 & 3.94E-08 & 0.85(L)  \\

Congestion Duration (s) & 
3.7 -- 3.77 & 0.2308 & 0.59 (S) & 
3.73 -- 5.5 & 0.0005 & 0.76 (L) & 
2.9 -- 3.43 & 0.0037 & 0.71(M)  \\ 

Packet Loss (\%)  & 
0.03 -- 25.38 & 2.065E-11 & 1 (L) & 
 14.33 -- 17.88 & 6.108E-10 & 0.97(L) & 
0.03 -- 24.56 & 1.893e-11 & 1(L)   \\

Exec Time (ms) & 
550.8 -- 1006.84 & < 2.2E-16 & 0.88 (L) & 
1237.16 -- 2927.74 & < 2.2E-16 & 0.93(L) & 
588.86 -- 2594.24 & < 2.2E-16 & 1(L) \\
\bottomrule
\bottomrule
&
\multicolumn{3}{c||}{ \textbf{\emph{FULL(32,3)}}} & 
\multicolumn{3}{c||}{ \textbf{\emph{FULL(32,4)}}} & 
\multicolumn{3}{c||}{ \textbf{\emph{MNP(5,2)}}} 
	  \\
\hline
& \textbf{avg(G-D)}& \textbf{p-value} & \textbf{$\bf{\hat{A}_{12}}$} &
 \textbf{avg(G-D)}& \textbf{p-value} & \textbf{$\bf{\hat{A}_{12}}$} &
 \textbf{avg(G-D)}& \textbf{p-value} & \textbf{$\bf{\hat{A}_{12}}$}  \\  
\hline
\# Congestion & 
 3.7 -- 5  & 0.0065  & 0.68(M) & 
 3 -- 3.03 & 0.8456 & 0.51(N) & 
1.8 -- 2.2 & 0.002 & 0.68(M)  \\

Congestion Duration (s) & 
5 -- 5.57 & 0.0647 & 0.64(S) & 
3.73 -- 3.33 & 0.1643 & 0.40(S) & 
3.87 -- 5.23 & 0.0029  & 0.71(M)   \\ 

Packet Loss (\%)  & 
14.16 -- 20.93 & 1.382E-06 & 0.86(L) & 
20.14 -- 24.85 & 0.0160 & 0.68(M) & 
32.39 -- 32.65  &   0.004 & 0.72(M)  \\

Exec Time (ms) & 
2654.62 -- 3230.97 & < 2.2E-16 & 0.80(L) & 
2756.13 --  2967.30 & 8.188E-12 & 0.79(L) & 
381.07 -- 469.13 & 0.099 & 0.59(S)  \\
\bottomrule
\bottomrule
  &
\multicolumn{3}{c||}{ \textbf{\emph{MNP(8,2)}}} & 
\multicolumn{3}{c||}{ \textbf{\emph{MNP(12,2)}}} & 
\multicolumn{3}{c||}{ \textbf{\emph{MNP(17,2)}}} 
\\
\hline
& 
 \textbf{avg(G-D)}& \textbf{p-value} & \textbf{$\bf{\hat{A}_{12}}$} &
 \textbf{avg(G-D)}& \textbf{p-value} & \textbf{$\bf{\hat{A}_{12}}$} &
 \textbf{avg(G-D)}& \textbf{p-value} & \textbf{$\bf{\hat{A}_{12}}$} \\  
\hline
\# Congestion & 
2.1 -- 3.4 & 1.55E-08 & 0.9(L) &
2.6 -- 4.03 & 2.985E-09 & 0.89(L) & 
2.43 -- 5.7 & 1.479E-11 & 0.99(L) \\

Congestion Duration (s) & 
5.57 -- 7.6 &  0.0002  &   0.77(L) & 
3.53--4.33 &  0.0090  &  0.68(M)  & 
3.6 -- 6.67 &  5.509E-07  &  0.87(L) \\ 

Packet Loss (\%)  & 
32.6 -- 33.6    &   3.43E-06 & 0.85(L) &
18.24 -- 20.73  &  0.0029  & 0.72(M) & 
13.98 --  16.52 & 0.1153  &  0.62(S)  \\

Exec Time (ms) & 
787.06 -- 392.62 & <2.2E-16 & 0.08(L) & 
468.97 -- 370.20 & 0.4623 & 0.53(N) & 
485.26 -- 619.83 &1.28E-12 & 0.79(L)\\
\bottomrule
\end{tabular}}

\vspace*{.1cm}	
{\scriptsize 
\hspace{.5cm} * The avg(G-D) column shows the average (avg) metrics for GenAdapt (G) versus those for DICES (D).\hfill\mbox{}}
\end{table}

With the analysis for RQ1 intuitively explained over one subject, namely MNP(8, 2), we now present the complete results for this RQ. Table~\ref{tab:GpvsDices} compares GenAdapt against DICES for our eighteen synthetic subjects (i.e., seven FULL($n$, 3), seven FULL ($n$, 4) and four MNP($n$, 2) where $n$ is the number of network nodes) by reporting the average number  of congestion occurrences, the average total duration that the network is congested, the average execution time, and the average packet-loss values obtained by 30 runs of DICES and GenAdapt. For four subjects, some runs of DICES failed to resolve the last congestion within the simulation time. Specifically, for FULL(6, 3), 12 runs; for FULL(6, 4), 10 runs; for FULL(7, 3), 20 runs; and for FULL(10, 4), 3 runs of DICES failed to resolve congestion. For these subjects, the reported average number of congestion occurrences and execution times capture only the successful runs of DICES. In contrast, all runs of GenAdapt for all the subjects successfully resolved congestion within the simulation time.

We compare the results of Table~\ref{tab:GpvsDices} through statistical testing. We use the non-parametric pairwise Wilcoxon rank sum test~\cite{capon:91} and the Vargha-Delaney's $\hat{A}_{12}$ effect size~\cite{vargha:00}.
We first focus on the following three metrics: number of congestion occurrences, congestion duration, and packet loss. For nine subjects -- MNP(12, 2), MNP(8, 2), MNP(5, 2), FULL(22, 4), FULL(22, 3), FULL(15, 3), FULL(10, 4), FULL(6, 4), and FULL(6, 3) -- the $p$-values for all the comparisons of the above three metrics are lower than $0.05$ and the $\hat{A}_{12}$ statistics show large or medium effect sizes, indicating that GenAdapt significantly improves over DICES with respect to the three metrics on the nine subjects. For the seven other subjects -- \hbox{FULL(5, 3)}, FULL(7, 3), FULL(7, 4), FULL(10, 3), FULL(15, 4), FULL(32, 3) and MNP(17, 2) -- GenAdapt significantly improves over DICES with large or medium effect sizes with respect to at least one of these three metrics. For all the three metrics of the all the subjects except for the congestion duration metric of FULL(32, 4), the averages obtained by GenAdapt are better than those obtained by DICES. For FULL(32, 4), we note that while the average congestion duration  for GenAdapt is slightly lower than that for DICES, the statistical test shows no difference between the congestion duration values obtained from DICES and GenAdapt (i.e., the $p$-value is higher than 0.05). We observe that, for FULL(32, 4), the packet loss and the execution time of GenAdapt are significantly better than those of DICES.

\begin{figure}[t]
	\centerline{\includegraphics[width=0.5\columnwidth]{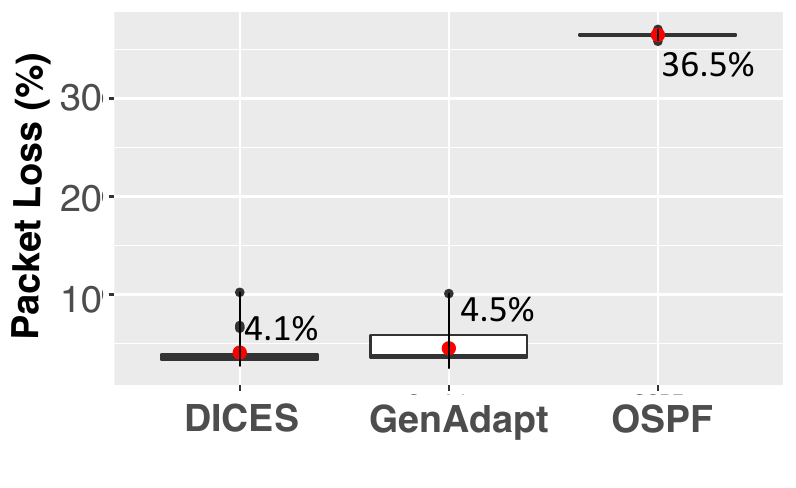}}
	\caption{Comparing packet-loss values obtained by 30 runs of DICES, GenAdapt and OSPF.}\label{fig:EMSeval}

\end{figure}

 Figure~\ref{fig:EMSeval} shows the packet-loss values obtained by 30 runs of GenAdapt, DICES and OSPF applied to the industry subject (EMS). All  30 runs of GenAdapt and DICES could resolve the congestion in EMS within the simulation time. However, OSPF incurs high packet loss (avg. $36.5$\%), since its heuristic link weights cannot  prevent congestion in EMS. Both GenAdapt and DICES start with high packet loss, but since they can resolve the congestion, packet loss drops quickly, yielding  low averages over the duration of simulation. The differences between the packet-loss values of  GenAdapt and DICES are neither statistically significant ($p$-value = 0.44) nor practically significant -- a packet-loss difference of 0.4\%  is negligible in practice.  The results of Figure~\ref{fig:EMSeval} signify the need for runtime adaptation in EMS. Further, the results in both Table~\ref{tab:GpvsDices} and Figure~\ref{fig:EMSeval}  show that while GenAdapt can reduce congestion occurrences over time compared to DICES, doing so does not come at the cost of being less effective in resolving a one-off congestion in EMS caused by several requests arriving almost at once.

\vspace*{.15cm}
 \resq{The answer to \textbf{RQ1} is that GenAdapt successfully resolves congestion in all the subjects. Compared to DICES, GenAdapt reduces the average number of congestion for thirteen subjects with a high statistical significance, while DICES never outperforms GenAdapt in any of the congestion resolution metrics. Further, for our industry subject, GenAdapt significantly outperforms the standard SDN data-forwarding algorithm, OSPF, in reducing packet loss.}

\subsection{RQ2 -- Transferability}
\label{sec:rq1}
Recall that in its first invocation (i.e., the first time self-adaption is needed), GenAdapt uses a randomly generated initial population, and only in its subsequent invocations, it reuses in its initial population half of the best solutions computed previously. To answer RQ2, we modify the GenAdapt configuration used in RQ1 as follows: For each of the FULL and MNP graphs, we designate the smallest subject as \emph{BaseSubject}. Specifically, FULL(5, 3), FULL(5, 4) and  MNP(5, 2) are BaseSubjects for the FULL(\_, 3), FULL(\_, 4) and MNP(\_, 2) categories, respectively. For each category,  we use half of the best solutions (ordered based on their fitness values) obtained from the BaseSubject of that category to initialize the population of GenAdapt for the two largest graphs in that category. More precisely, the best solutions obtained from  FULL(5, 3) are reused to initialize the populations for FULL(22, 3) and FULL(32, 3), the best solutions obtained from  FULL(5, 4) are reused to initialize the populations for FULL(22, 4) and FULL(32, 4), and the best solutions obtained from  MNP(5, 2) are reused to initialize the populations for MNP(12, 2) and MNP(17, 2).

Table~\ref{tab:GpFSvsGp} shows the results for the two largest networks in each subject category (i.e., FULL(22, 3), FULL(32, 3), FULL(22, 4), FULL(32, 4), MNP(12, 2) and MNP(17, 2)) obtained from the modified GenAdapt, as described above, against the results obtained from the original GenAdapt configuration described in RQ1. We refer to the modified GenAdapt as GenAdapt\_Reuse and abbreviate it to GR in Table~\ref{tab:GpFSvsGp}; the original GenAdapt algorithm is abbreviated to G in the table.

We report in Table~\ref{tab:GpFSvsGp} the same four metrics as introduced in RQ1. Like in RQ1, we first consider the following three metrics: number of congestion occurrences, congestion duration, and packet loss. In all the comparisons in Table~\ref{tab:GpFSvsGp}, the averages of these three metrics obtained by GenAdapt\_Reuse are lower than those obtained by GenAdapt with the exception of the packet loss for FULL(22,4), where the averages for GenAdapt\_Reuse and GenAdapt are the same. For five out of six subjects - FULL(22, 3), FULL(32, 3), FULL(32, 4), MNP(12, 2) and MNP(17, 2) - the $p$-values for at least one of the above three metrics are lower than $0.05$ and the $\hat{A}_{12}$ statistics show large or medium effect sizes, indicating that GenAdapt\_Reuse significantly improves GenAdapt with respect to at least one of these three metrics on the five subjects. Notably, for the two MNP cases, GenAdapt\_Reuse significantly outperforms GenAdapt for all three metrics  with a large or medium effective size.

\begin{table*}[th]
	\caption{Statistical-test results comparing the average number of congestion occurrences, the duration of congestion, the execution-time and the packet-loss values obtained by 30 runs of GenAdapt\_Reuse (GR) versus GenAdapt (G).}
	\label{tab:GpFSvsGp}
	\centering
	\scalebox{0.6}{
	\begin{tabular}{||c|c|c|c||c|c|c||c|c|c||}
	\toprule
  & \multicolumn{3}{c||}{ \textbf{\emph{FULL(22,3)}}} & 
	  \multicolumn{3}{c||}{ \textbf{\emph{FULL(22,4)}}} & 
	  \multicolumn{3}{c||}{ \textbf{\emph{FULL(32,3)}}} 
	  \\
\hline
& \textbf{avg(GR-G)}& \textbf{p-value} & \textbf{$\bf{\hat{A}_{12}}$} &
 \textbf{avg(GR-G)}& \textbf{p-value} & \textbf{$\bf{\hat{A}_{12}}$} &
 \textbf{avg(GR-G)}& \textbf{p-value} & \textbf{$\bf{\hat{A}_{12}}$}  \\  
\hline
\# Congestion & 
 1.43 --2.1  & 0.0535 & 0.63(S) & 
 2.16 -- 2.3 & 0.3312 & 0.55(N) & 
 2.4 -- 3.7  & 0.0126 & 0.68(M)  \\

Congestion Duration (s) & 
1.63 -- 3.73 &  0.0001 & 0.78(L) & 
2.63 -- 2.9 & 0.2251 & 0.58(S) & 
3.2 -- 5 & 0.0480 & 0.65(S) \\ 

Packet Loss (\%)  & 
12.09  -- 14.33 & 3.248E-07 & 0.88(L) & 
0.03 -- 0.03 & 0.8531 & 0.51(N) & 
12.09 -- 14.16 & 0.6467 & 0.47(N) \\

Exec Time (ms) & 
1468.23 -- 1237.16 &  0.0008 & 0.31(M) & 
808.92 -- 588.86 & 4.601E-13 & 0.14(L) & 
2017.63.04 -- 2654.62 & 0.0454 & 0.59(S)  \\
\bottomrule
\bottomrule
  & 
	  \multicolumn{3}{c||}{ \textbf{\emph{FULL(32,4)}}}&  
	  \multicolumn{3}{c||}{ \textbf{\emph{MNP(12,2)}}} & 
	  \multicolumn{3}{c||}{ \textbf{\emph{MNP(17,2)}}} 
	  \\
\hline
& \textbf{avg(GR-G)}& \textbf{p-value} & \textbf{$\bf{\hat{A}_{12}}$} &
 \textbf{avg(GR-G)}& \textbf{p-value} & \textbf{$\bf{\hat{A}_{12}}$} &
  \textbf{avg(GR-G)}& \textbf{p-value} & \textbf{$\bf{\hat{A}_{12}}$}  \\  
\hline
\# Congestion & 
2.6 --  3 & 0.223 & 0.59(S) & 
1.1 -- 2.6 & 1.819E-11 & 0.97(L) & 
1.43 -- 2.43 & 4.941E-06 & 0.82(L) \\

Congestion Duration (s) & 
3.57 -- 3.73 & 0.844 & 0.52(N) & 
1.5 -- 3.53 & 7.106E-08 & 0.89(L)  & 
1.87 -- 3.6 & 3.096E-05 & 0.81(L)  \\ 

Packet Loss (\%)  & 
3.69 -- 20.14 & 4.271E-06 & 0.85(L) & 
16.15 -- 18.24 & 0.0004 & 0.77(L) & 
12.56 --  13.98 & 0.0127 & 0.69(M) \\

Exec Time (ms) & 
1989.06 -- 2756.13 & 0.0097 & 0.62(S) & 
229.61 -- 468.97 & 0.0004 & 0.71(M) & 
498.05 -- 485.26 & 0.0798 & 0.60(S) \\
\bottomrule
	\end{tabular}}

\vspace*{.1cm}	
{\scriptsize 
\hspace{.5cm} * The avg(GR-G) column shows the average (avg) metrics for GR versus G.\hfill\mbox{}}
\end{table*}

In relation to execution time, we observe from  Table~\ref{tab:GpFSvsGp} that the average execution time of GenAdapt\_Reuse is sometimes higher than that of GenAdapt. This increase in execution time is likely attributable to the fact that some of the link-weight functions ranked as best solutions and reused in the initial population may have complex and redundant structures. This causes an overhead in the execution time of GenAdapt\_Reuse. We note that the execution time metric refers to the average execution time of a single invocation of a self-adaptation technique. The number of times a self-adaptation technique is required to be called to solve a congestion  is not included in this metric, but is represented through the number of congestion occurrences and the total congestion duration metrics. In all the cases where the average of a single execution time of GenAdapt\_Reuse takes longer than that of GenAdapt (i.e., FULL(22, 3), FULL(22, 4), MNP(17, 2)), GenAdapt\_Reuse still outperforms GenAdapt with respect to the number of congestion occurrences and total congestion duration. This shows that GenAdapt\_Reuse  is overall more effective than GenAdapt in resolving congestion occurrences observed over a time period.

In addition, we compare in Table ~\ref{tab:GpFSvsDices} the results from  GenAdapt\_Reuse against DICES and perform statistical tests for our four metrics. For five synthetic subjects - FULL(22, 3), FULL(22, 3), FULL(32, 3), MNP(12, 2) and MNP(17, 2)- GenAdapt\_Reuse significantly outperforms DICES with respect to  all the four metrics. For FULL(32, 4), GenAdapt\_Reuse significantly improves over DICES for packet loss and  execution time, and for the two other metrics, i.e., number of congestion occurrences and congestion duration, GenAdapt\_Reuse yields higher averages than DICES.

\begin{table*}[th]
	\caption{Statistical-test results comparing the average number of congestion occurrences, the duration of congestion, the execution-time and the packet-loss values obtained by 30 runs of GenAdapt\_Reuse (GR) versus Dices (D).}
	\label{tab:GpFSvsDices}
	\centering
	\scalebox{0.6}{
	\begin{tabular}{||c|c|c|c||c|c|c||c|c|c||}
	\toprule
  & \multicolumn{3}{c||}{ \textbf{\emph{FULL(22,3)}}} & 
	  \multicolumn{3}{c||}{ \textbf{\emph{FULL(22,4)}}} & 
	  \multicolumn{3}{c||}{ \textbf{\emph{FULL(32,3)}}} 
	  \\
\hline
& \textbf{avg(GR-D)}& \textbf{p-value} & \textbf{$\bf{\hat{A}_{12}}$} &
 \textbf{avg(GR-D)}& \textbf{p-value} & \textbf{$\bf{\hat{A}_{12}}$} &
 \textbf{avg(GR-D)}& \textbf{p-value} & \textbf{$\bf{\hat{A}_{12}}$}  \\  
\hline
\# Congestion & 
 1.43 -- 5 & 2.57E-13 & 1(L) & 
 2.16 -- 3 & 8.986E-11 & 0.92(L) & 
 2.4 --  5 &  2.054E-09 & 0.92(L)  \\

Congestion Duration (s) & 
1.63 -- 5.5 & 7.248E-11  & 0.97(L) & 
2.63 -- 3.43 & 1.546E-05 & 0.81(L) & 
3.2 -- 5.57 & 8.265E-06 & 0.83(L) \\ 

Packet Loss (\%)  & 
12.09  -- 17.88 & < 2.2E-16 & 1(L) & 
0.03 -- 24.56 & 1.956E-11 & 1(L) & 
12.09 -- 20.93 & 3.33E-11 & 1(L) \\

Exec Time (ms) & 
1468.23 -- 2927.74 & < 2.2E-16  & 0.95(L) & 
808.92 -- 2594.24 & < 2.2E-16 & 1(L) & 
2017.63.04 -- 3230.97 & < 2.2E-16 & 0.90(L)  \\
\bottomrule
\bottomrule
  & 
	  \multicolumn{3}{c||}{ \textbf{\emph{FULL(32,4)}}}&  
	  \multicolumn{3}{c||}{ \textbf{\emph{MNP(12,2)}}} & 
	  \multicolumn{3}{c||}{ \textbf{\emph{MNP(17,2)}}} 
	  \\
\hline
& \textbf{avg(GR-D)}& \textbf{p-value} & \textbf{$\bf{\hat{A}_{12}}$} &
 \textbf{avg(GR-D)}& \textbf{p-value} & \textbf{$\bf{\hat{A}_{12}}$} &
  \textbf{avg(GR-D)}& \textbf{p-value} & \textbf{$\bf{\hat{A}_{12}}$}  \\  
\hline
\# Congestion & 
2.6 -- 3.03  & 0.0650 & 0.61(S) & 
1.1 -- 4.03 & 9.448E-14 & 1(L) & 
1.43 -- 5.7 & 5.247E-12 & 1(L) \\

Congestion Duration (s) & 
3.57 -- 3.33 & 0.2156 & 0.41(S) & 
1.5 -- 4.33 & 1.404E-12 & 1(L)  & 
1.87 -- 6.67 & 1.311E-10 &  0.98(L) \\ 

Packet Loss (\%)  & 
3.69 -- 24.85 & 1.362E-08 & 0.93(L) & 
16.15 -- 20.73 & 4.19E-10 & 0.97(L) & 
12.56 -- 16.52  &  0.0083 & 0.70(M) \\

Exec Time (ms) & 
1989.06 -- 2967.30 & 7.102E-13 & 0.82(L) & 
229.61 -- 370.20 & 7.76E-12 & 0.89(L) & 
498.05 -- 619.83 & 8.675E-11 & 0.82(L) \\
\bottomrule
	\end{tabular}}

\vspace*{.1cm}	
{\scriptsize 
\hspace{.5cm} * The avg(GR-D) column shows the average (avg) metrics for GR versus D.\hfill\mbox{}}
\end{table*}

\vspace*{.15cm}
\resq{The answer to \textbf{RQ2} is that bootstrapping GenAdapt with the best solutions it computes on a smaller network improves the algorithm's effectiveness for larger networks with the same topology. In particular, in all our study subjects, bootstrapping reduced either the average number of congestion occurrences or packet loss (or both). For the MNP topology, reductions in the average number of congestion occurrences and packet loss 
are statically significant.}
\vspace*{-.15cm}

\subsection{RQ3 -- Scalability}
To answer RQ3, we perform two sets of studies. In the first set, we increase the network size and in the second set -- the number of requests. Specifically, first, we create ten synthetic networks with the FULL topology and having  5, 10, \ldots , 50 nodes; and, for each network, we generate five requests with the same bandwidth at once to reach a total bandwidth of 150Mbps.  Second, we create a five-node network  with the FULL topology and execute ten different experiments by subsequently generating  5, 10, \ldots , 50 requests at once such that for each experiment, the requests have the same bandwidth and the sum of the requests' bandwidths is 150Mbps. For both experimental sets, we record the execution time of the configuration of GenAdapt used for the FULL topology as described in Section~\ref{sec:rq1}. Our experimental setup in RQ3 follows that used by DICES for scalability analysis. We focus on the FULL topology for scalability analysis because networks with this topology have considerably more links and pose a bigger challenge for self-adaptation planning, as evidenced by the failure of  DICES over several networks with the FULL topology (see Table~\ref{tab:GpvsDices}).

\begin{figure}[t]
	\centerline{\includegraphics[width=0.5\columnwidth]{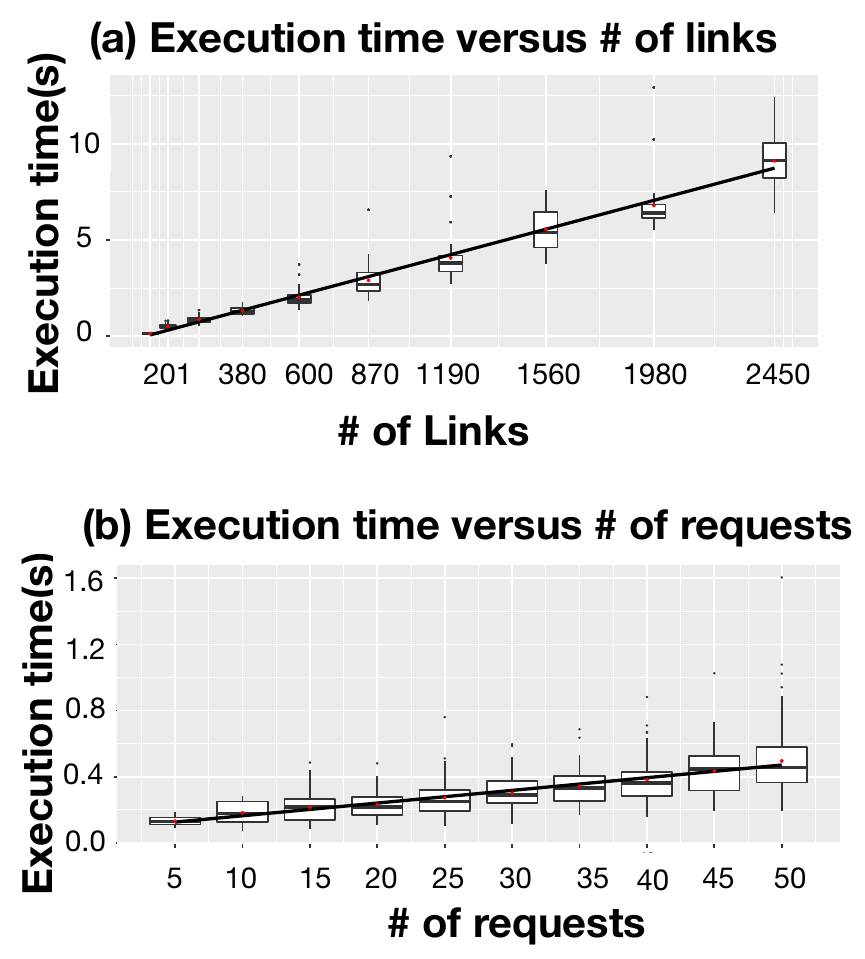}}
	\caption{Execution times of GenAdapt versus (a) number  of links and (b) number of requests.}
	\label{fig:rq2}
\end{figure}

Figure~\ref{fig:rq2} shows the execution time of GenAdapt versus the number of network links (Figures~\ref{fig:rq2}(a)), and  versus the number of requests (Figure~\ref{fig:rq2}(b)). For each network size and for each number of requests, GenAdapt is executed 30 times. For each diagram in Figures~\ref{fig:rq2}, we have fitted a linear regression line ($\mathit{time} = -0.019 + 0.0036 \times \mathit{links}$ in Figure~\ref{fig:rq2}(a); and $\mathit{time} = 0.087  + 0.0077 \times \mathit{requests}$ in Figure~\ref{fig:rq2}(b)). In both cases, we obtain a $p$-value of $< 2.2e-16$, indicating that the models fit the data well.

\vspace*{.15cm}
\resq{The answer to \textbf{RQ3} is that the execution time of GenAdapt is linear in the number of requests and in the size of the network.}
\vspace*{-.15cm}

\subsection{Threats to Validity}
Internal, construct and external validity  are the validity aspects most pertinent to our evaluation.

\textbf{Internal validity.} In our approach, we treat the threshold that determines network congestion as a static and fixed parameter. A fixed threshold may not be suitable for all situations as it assumes that network congestion will always occur when a certain threshold is exceeded. This may lead to false alarms or missed opportunities for adaptation as network conditions can vary and different factors may contribute to congestion.
A more flexible approach, such as proactively predicting network congestion using machine learning, may be more effective in certain situations. Changing the fixed threshold parameter of an SDN to a dynamic parameter would not affect our approach in the planning step, which is our main contribution. Specifically, the grammar of the link-weight functions would remain unchanged, and the only difference would be the use of a dynamic threshold obtained from the analyzing step instead of a static one.

\textbf{Construct validity.} Our experiments are based on a testbed; the extent to which the measurements obtained through the testbed are reflective of the real world is therefore an important factor to consider. We note that our testbed uses ONOS, an actual \emph{carrier-grade} SDN controller, which can be used with real networks as well~\cite{DBLP:journals/csur/ZhuKSXLDG21,DBLP:conf/cnsm/AzzouniBP17,Berde:14,Bianco:17}.  The network emulator that our testbed builds on, i.e., Mininet, is widely considered to be a \emph{high-fidelity} network emulator.

\textbf{External validity.} 
The centralized architecture adopted in our work may face challenges if deployed in very large networks. We note three ways to mitigate  potential scalability issues: 
(1)~Transferring to a larger network the best control logic learned on a smaller network that has the same topology as the larger one. As demonstrated by RQ2, GenAdapt's performance improves for a larger network by bootstrapping GenAdapt with best solutions computed on a smaller network with the same topology.  
(2)~Localizing congestion to a subset of the network and applying GenAdapt to that subset. In this case, one would need to account for only the topology and traffic information of the subset, thus improving scalability. 
(3)~Deploying the adaptation component on a cluster, as commonly done in service-oriented architectures~\cite{Pahl:18}. By doing so, GenAdapt can leverage more computational resources, which may be necessary when dealing with very large networks.
At the same time, we note that, based on the results of RQ3, our approach is scalable to networks with up to 2500 links, which corresponds to a full network with 50 nodes. The number, size and complexity of the networks in our experiments are comparable to those in the literature, e.g.,~\cite{ShinNSB0Z20,DBLP:conf/cnsm/AzzouniBP17,DBLP:conf/cscwd/LiuZGQ18,DBLP:journals/csur/ZhuKSXLDG21,Agarwal:13}. 

We scoped our experiments to two network topologies only. As discussed in Section~\ref{sec:rq1}, the two topologies are the two extreme ends of the spectrum in terms of path overlaps between node pairs -- the main determinant of complexity for GenAdapt. We expect that, for other topologies lending themselves  to congestion resolution via re-routing, our approach will behave within the same range as seen in our experiments. To ensure adequate coverage, our evaluation examined 26 networks based on the two topologies considered. In addition, our evaluation included a real industrial IoT network with its own topology. The number, size and complexity of the networks in our experiments are comparable to those in the literature, e.g.,~\cite{ShinNSB0Z20,DBLP:conf/cnsm/AzzouniBP17,DBLP:conf/cscwd/LiuZGQ18,DBLP:journals/csur/ZhuKSXLDG21,Agarwal:13}. 
To show the effectiveness of our approach (i.e., RQ1), we present  results from 18 synthetic subjects. In comparison, the paper that introduces DICES~\cite{ShinNSB0Z20} uses only one synthetic subject to show effectiveness, and the earlier version of this work~\cite{Li:22} uses ten synthetic subjects. As an additional external validity measure for effectiveness, we examine (through RQ2) the reuse of adaptation solutions across networks and demonstrate that such reuse improves effectiveness.
A different consideration related to external validity is the tuning of our approach. The only parameter of our approach that requires tuning by the user is the number of generations of GenAdapt. As discussed in Section~\ref{sec:rq1}, tuning this parameter is straightforward based on the easily measurable objective of keeping the execution time of GenAdapt less than $\delta$ (i.e., the time interval between consecutive invocations of GenAdapt). In practice, engineers can use our simulator to tune this parameter for their specific network topology and traffic profile.


\section{Related Work}
\label{sec:related}

We compare our approach with the related work on self-adaptation frameworks, learning-based self-adaptation, and  congestion control in SDNs.

\subsection{Self-adaptive Systems}
To better position our work within the software engineering literature on self-adaptive systems, we present in
Table~\ref{tab:relatedwork} a structured comparison with the research strands most closely related to our work. Our comparison takes the following  criteria into consideration:

\begin{table}
\caption{
Comparison of GenAdapt with relevant work strands in the literature on self-adaptive systems.}
\label{tab:relatedwork}
\vspace*{-.5em}
\begin{center}
\scalebox{.8}{
\begin{tabular}{|p{2cm} |p{1.8cm} |p{1.8cm} | p{1.5cm}|p{1.7cm}|p{1.9cm}|p{1.4cm} | p{1.5cm} |}
\toprule
\multirow{2}{3em}{Approach(es)} & \multirow{2}{4em}{Adaptation Paradigm} & \multirow{2}{5em}{Area of Autonomic Computing} & \multirow{2}{4em}{Application Domain} &  \multicolumn{3}{c|}{Planning Strategy}\\[2ex]
&&&& \multicolumn{1}{c|}{Online/Offline} & \multicolumn{1}{c|}{Generative} & \multicolumn{1}{c|}{Transferable} \\[1em]
 \hline
\cite{Kramer:07}&Architecture-based&General&Generic&Offline&$\times$&$\times$
 \\
 \hline
\cite{Garlan:04}\cite{Oreizy:99}&Architecture-based&General&Generic&Unspecified&$\times$&$\times$
 \\
  \hline
\cite{Inverardi:10}&Requirement-based&General&Generic&Unspecified&$\times$&$\times$
 \\
   \hline
\cite{Morin:08, Morin:09}&Model-based&General&Generic&Unspecified&$\times$&$\times$
 \\
    \hline
\cite{Bhuiyan:17}&Architecture-based&Self-optimizing&Network&Offline&$\times$&$\times$
 \\
 \hline
\cite{Stein:16} &Model-based&Self-configuring&Network&Offline&$\times$&$\times$
 \\
 \hline
 \cite{Anaya:14}&Model-based&Self-configuring&Network&Offline&$\times$&$\times$
 \\
\hline
\cite{ShinNSB0Z20}&Learning-based&Self-healing&IoT&Online&$\times$ &$\times$
 \\
 \hline
\cite{Elkhodary:10}\cite{Rodrigues:18}&Learning-based&Self-configuring&Generic&Online&$\times$ &$\times$
\\
\hline
\cite{Camara:20}&Learning-based&Self-configuring&IoT&Online&$\times$ &$\times$
\\
\hline
\cite{Weyns:22}\cite{Quin:19}&Learning-based&Self-configuring&IoT&Online&$\times^*$ &$\times$
\\
\hline
\cite{Jamshidi:19}&Learning-based&Self-configuring&Robotics&Online&$\times$&$\times$
\\
\hline
GenAdapt&Learning-based&Self-healing&IoT&Online&$\checkmark$&$\checkmark$
\\
\bottomrule
\end{tabular}
}
\end{center}
\raggedright\footnotesize{\it * These approaches use Deep Learning (DL). DL can be used for generative tasks as well. However, in the existing self-adaptation literature, DL has been used mainly for classification and regression use cases.}
\vspace*{-.35cm}
\end{table}

(1)~\emph{Adaptation paradigm} refers to the method used to engineer self-adaptive systems~\cite{Weyns2021}.  In alignment with Weyns' classifications of adaptation paradigms (or waves)~\cite{Weyns2021}, we categorize related research as \emph{architecture-based}, \emph{model-based}, \hbox{\emph{requirement-based}, \emph{control-based}, or \emph{learning-based}.}

(2)~\emph{Autonomic computing area} is a classification  introduced by IBM in the early 2000s~\cite{IBM:05},  inspired by the human body's autonomic system. It refers to the self-managing characteristics of distributed computing resources, adapting to unpredictable changes while hiding intrinsic complexity to operators and users. The main autonomic computing areas are: \emph{self-configuring}, \emph{self-healing}, \emph{self-optimizing} and \emph{self-protecting}. Briefly,  self-configuring refers to the ability of the system to dynamically adapt to changes in the environment; self-healing describes the capability to discover, diagnose and act to prevent disruptions; self-optimizing entails the ability to optimize the system's resource utilization; and self-protecting involves anticipating, detecting, identifying and protecting against hostile behaviours. 
While some approaches are specific to  a particular area of adaptation, some others propose architectures or models that can benefit several or all areas of autonomic computing. These latter approaches are labeled  ``General'' in Table~\ref{tab:relatedwork}.

(3)~\emph{Application domain} refers to the area or field where a given approach is applied. As shown in Table~\ref{tab:relatedwork}, some approaches are ``Generic'' and not tied to any particular domain. The more recent approaches, e.g., \cite{Bhuiyan:17, Stein:16, Anaya:14, ShinNSB0Z20, Camara:20, Quin:19, Weyns:22, Jamshidi:19}, are typically targeted at  specific domains.

(4)~\emph{Planning strategy}  relates to the planning step of the MAPE-K loop (Figure~\ref{fig:fmw}) and characterizes the strategy used in this step according to three factors:  (i) whether the strategy is online or offline, (ii) whether the planning is generative, and (iii) whether the results of planning are transferable.
We distinguish between online and offline planning as follows: Online planning strategies continuously learn and adapt while in operation, adjusting to new system behaviour ``on the fly'', while offline planning strategies lack the ability to dynamically learn and apply at runtime new knowledge gained during system execution.  Offline strategies either employ pre-designed static policies or involve a distinct pre-processing training phase that is  not updated based on new data from the system in operation. A generative planning approach is designed to update complex structures or formulas used by a system, enabling the system to produce effective planning solutions at runtime and  reducing the need for the invocation of the planning step. A transferable approach generalizes solutions obtained over one system to other systems that share certain common characteristics.
Our contribution enhances the planning step of self-adaptation by making it generative and transferable.

Having set the stage with Table~\ref{tab:relatedwork}, we now discuss the specific work strands listed in the table. We begin with an overview of self-adaptation frameworks, followed by an outline  of more recent developments where  learning is used for self-adaptation.

\textbf{Self-adaptation frameworks.} 
Engineering self-adaptive systems, including the principles underlying the construction, maintenance and evolution of such systems, have been studied from different angles and for different domains~\cite{Coker:15,Bhuiyan:17,Anaya:14,Stein:16,DBLP:books/daglib/p/GarlanSC09,ChengRM13}. 
Kramer and Magee~\cite{Kramer:07} propose a layered architecture model (component control, change management, and goal management) as an  abstraction mechanism for modelling and reasoning about dynamic adaptations. 
Garlan et al.~\cite{Garlan:04} develop the Rainbow framework -- a reusable architecture-based framework for self-adaptive systems. Rainbow uses a rule-based language to implement self-adaptation rules, with the Rainbow architecture model monitoring and detecting the need for adaptation  in a managed system.

Inverardi and Mori~\cite{Inverardi:10} propose a feature-based self-adaptation framework to tackle the evolution of requirements for high-variability systems. A \emph{feature} is a dynamic unit
representing the smallest part of a service that the user can recognize. The proposed framework handles the consistent evolution of component-based service-oriented systems by adapting their features. 
Oreizy et al.~\cite{Oreizy:99} propose the C2 architectural style to minimize inter-dependencies between components by communicating messages between them through connectors. The proposed architecture is used as a basis for architecture-based runtime adaptation and evolution.
Morin et al.~\cite{Morin:08, Morin:09} propose a self-adaptation approach based on aspect-oriented modelling. This approach uses aspects as course-grained adaptation units to avoid combinatorial explosion in the adaptation space. The approach employs feature models to capture  variability in the system and its context. The combination of aspects and feature models leads to a versatile platform for realizing self-adaption.

Bhuiyan et al.~\cite{Bhuiyan:17} propose an approach for autonomous adaptive sampling for network systems. The adaptive rate sampling is implemented as an embedded algorithm and evaluated using simulators . Stein et al.~\cite{Stein:16} present a network topology adaptation model and rule language, and demonstrate how the model and the adaptation rule language can be employed in a self-adaptation monitoring loop. In this work, network links are updated or removed in order to adapt to environmental changes. Anaya et al.~\cite{Anaya:14} propose a framework for proactive adaptation,  relying on predictive models and historical information about the environment. The predictive models are specifically used for improving the decisions generated by the reasoning engine of the framework.

The above  approaches aim to improve the engineering of self-adaptive systems by proposing new architectures or by utilizing advancements in  software modelling and requirements engineering. In our work, in order to design a self-adaptive SDN framework,  we use a centralized architecture, drawing inspiration from the MAPE-K loop. In that respect,  our work has been informed by the above approaches, since  they too are largely based on MAPE-K. Movahedi et al.~\cite{Movahedi:12} classify the proposed architectures for network autonomy into centralized, distributed and hybrid architectures. 
While distributed and hybrid architectures may offer better scalability, flexibility and fault tolerance compared to centralized architectures~\cite{Movahedi:12},  they also introduce additional overhead. In the case of distributed architectures, this overhead is due to the communication required for achieving consensus among distributed adaptation managers. For hybrid architectures, the overhead arises from the need for optimal allocation of responsibilities and management tasks between the central and distributed entities.
Although we adopt a centralized architecture in our work, our contribution is not strictly tied to this decision. Our main contribution is in improving the planning step of self-adaptation, and this falls primarily under the umbrella of dynamic adaptive search-based software engineering~\cite{Harman:12}, which uses a blend of artificial intelligence and optimization for adaptation. In this context, the closest work to ours is DICES~\cite{ShinNSB0Z20}, which serves as our baseline and also employs a centralized SDN architecture. 
We  leave to future work the exploration of deploying  our   approach in  decentralized and hybrid architectures. We anticipate that such architectures can be supported if a comprehensive view of the network is available or if  congestion can be reliably localized to smaller subset(s) of a large network.

\textbf{Learning-based self-adaptation.} 
There is a growing interest in using machine learning techniques in self-adaptive systems~\cite{Saputri:20}.   FUSION~\cite{Elkhodary:10} uses a learning-based approach where, instead of relying on static analytical models, a machine-learning technique, named model trees learning (MTL), is used for configuring the adaptive behaviors of a system. Like our approach, FUSION aims to mitigate environment uncertainty by gradually learning suitable adaptation solutions as the system operates in a new environment. FUSION improves adaptation decisions using analytical models that predict the impact of such decisions on quantifiable features, e.g., system response.
Rodrigues et al.~\cite{Rodrigues:18} use decision trees to predict the time it takes to effect a change in the environment based on the current state of the system, and utilize this information to configure adaptation policies.
Camara et al.~\cite{Camara:20} use reinforcement learning and quantitative verification to reduce the decision space in a self-adaptation framework for IoT systems. Jamshidi et al.~\cite{Jamshidi:19}  provide an adaptation technique that, in an offline mode, learns a set of Pareto-optimal configurations, and then uses these configurations in adaptation plans at runtime. 
To reduce the adaptation space during the ``Analyze'' step of the MAPE-K loop, Quin et al.~\cite{Quin:19} propose a machine-learning framework based on classification and regression. In a similar manner, Weyns et al.~\cite{Weyns:22} employ deep learning for adaptation-space reduction, supporting three common classes of adaptation goals: threshold, optimization, and set-point goals.

Similar to the above approaches, we adopt a learning-based and online planning strategy for self-adaptation. Some of the above approaches use techniques such as decision trees to predict configuration parameters~\cite{Elkhodary:10, Rodrigues:18}, while others improve planning and decision making by narrowing the adaptation space at runtime using classification /  regression~\cite{Camara:20, Jamshidi:19, Quin:19}.   
Recently, Weyns et al.~\cite{Weyns:22} have used deep learning (DL) to reduce the adaptation space at runtime. DL can eliminate the necessity for domain engineers and further makes it possible to consider a combination of threshold, minimzation and set-point adaptation goals. To date and in the context of self-adpation, learning models have been used in a discriminative way, where learned models are applied to candidate adaptation options to identify a subset  satisfying the desired adaptation goals. In contrast, we employ a generative approach. Instead of using a prediction model to determine which link-weight values satisfy our adaptation goal, we use GP to generate a formula that, on demand,  produces link-weight values satisfying our goal.  In addition, we show that  our planning strategy is transferable. That is, the best formulas learned on smaller networks can be reused for larger networks, when both the smaller and larger networks have the same topology.

We note that DL models can be used in a generative way to produce  adaptation solutions, without applying the models to candidate solutions in the adaptation space. For the case studies used by Weyns et al.~\cite{Weyns:22}, the size of the adaptation space is a few thousand permutations, and hence, using DL models generatively was likely deemed unnecessary. In our work, however, the search space for link-weight values is much larger. 
We thus opt to use a generative approach, specifically genetic programming (GP), due to its flexibility in learning formulas that adhere to our specific grammar for link-weight functions. To our knowledge, we are the first to apply GP for adapting SDNs and reducing the need for frequent adaptations in this domain.

\subsection{Congestion control in SDNs} Network congestion resolution has been studied extensively~\cite{Mathis:96,Alizadeh:10,He:16,Betzler:16}.
Some congestion resolution approaches work by adjusting data transmission rates~\cite{Betzler:16}. Several others focus on traffic  engineering to provide solutions for better traffic control, better traffic operation, and better traffic management, e.g., by multi-path routing~\cite{DBLP:journals/wicomm/HanDGSW15},   creating a new routing technique~\cite{DBLP:conf/globecom/MaoFTKAIM17},  and flow-based routing~\cite{DBLP:conf/globecom/AmokraneLBP13}. Our work is related to the research that utilizes the additional flexibility offered by the programmable control in the SDN architecture~\cite{Brandt:16,Hong:13,Chiang:18,Agarwal:13,Gay:17,Huang:16,ShinNSB0Z20,Yashar:12}. Within this line of research, congestion control is commonly cast as an optimization problem~\cite{Chiang:18}, to be solved using local search~\cite{Gay:17} or linear programming~\cite{Agarwal:13}.
Furthermore, there are techniques, e.g., \cite{Yashar:12}, which use pre-defined rules, and techniques, e.g., \cite{Zuo:19}, which use machine learning (particularly deep learning) to dynamically mitigate SDN congestion at runtime.

Building on the OPPBG (OpenFlow-based path-planning with bandwidth guarantee) algorithm, Xiao et al.~\cite{Xiao:19} propose an SDN traffic-routing approach which computes best routing paths for new flows, taking into account both bandwidth requirements and the number of link hops. 
Zuo et al.~\cite{Zuo:19} use neural networks and build a sequence-to-sequence model to generate ideal routing paths. This approach uses attention mechanisms~\cite{Xu:15} to ensure reasonable order of the network nodes in a sequence and uses beam search to move out of local optima. 
Chiang et al.~\cite{Chiang:18} formulate multicast traffic routing in SDN as an optimization problem and propose an algorithm for computing efficient routing paths, considering bandwidth consumption, scalability and rerouting overhead.
\hbox{Gay et al.~\cite{Gay:17}} propose a minimal-time forwarding approach based on local search to lower link loads in SDN and thereby improve network response time to unexpected events such as significant traffic changes and network failures.
Ghobadi et al.~\cite{Yashar:12} propose OpenTCP -- a rule-based adaptation framework for resolving network congestion in SDN-based data centers. In this approach, when a congestion is detected, the SDN controller generates (through rules) a set of congestion-resolution actions, and distributes these actions to end hosts for execution.
Agarwal et al.~\cite{Agarwal:13} propose a graph-based algorithm for traffic engineering in cases where there is only a partial deployment of SDN capabilities and where SDNs co-exist with traditional networks. The authors show that network performance can be significantly improved even with a few \hbox{strategically deployed SDN switches.
}

In contrast to our approach, none of the  SDN congestion control techniques that we are aware of evolve the routing rules in response to feedback from SDN monitoring. Our approach is generative and its main purpose is, instead of directly predicting suitable solutions for adaptation, to learn a higher-order structure, i.e., a link-weight function, that generates optimal adaptations on demand. The generated link-weight function evolves the flow rules in the flow table by updating the link weights for incoming requests. While the routing algorithm itself is not changed, the flow rules are modified for the incoming requests to avoid future congestion. In this way,  our approach automatically learns forwarding rules that reduce the likelihood of future congestion occurrences and iteratively improves these rules when a congestion occurs.

\section{Conclusion}
\label{sec:conclusions}
We presented GenAdapt -- an adaptive approach that uses genetic programming for resolving network congestion in SDN-based IoT~networks. 
While existing self-adaptation research focuses on modifying a running system via producing individual and concrete elements, GenAdapt is generative and modifies the logic of the running system so that the system itself can generate the concrete elements without needing frequent adaptations. We used 18 synthetic and one industrial networks to compare GenAdapt against two baseline techniques:  DICES~\cite{ShinNSB0Z20}  and OSPF~\cite{Coltun:08,Cisco:05}. GenAdapt successfully resolved all congestion occurrences in our experimental networks, while DICES failed to do so in four of them. Further, compared to DICES, GenAdapt reduced the number of congestion occurrences and outperformed OSPF  in reducing packet loss. \textcolor{black}{Finally, our results show the transferability of the logic learned over networks that share the same topology but have different numbers of nodes. Specifically, we  observe that for a given topology, bootstrapping GenAdapt with the best logic learned on a smaller network can significantly improve performance on a larger
network in terms of the number of adaptation invocations, the amount of packet loss and the duration the network remains congested.}

An open challenge in self-adaptation is that evolutionary and statistical learning methods do not provide formal guarantees about the resolution of anomalies~\cite{DBLP:conf/seams/GheibiWQ21}. This challenge applies to GenAdapt as well. While GenAdapt had a 100\% success rate in our evaluation, our current approach does not guarantee that congestion will always be resolved when feasible. In the future, we would like to investigate ways to provide guarantees about congestion resolution through adaptation. 
Another area for future work is transitioning GenAdapt's design from a centralized architecture to a distributed or hybrid one in order to enable scaling to larger networks.  
Furthermore, we plan to expand our work to traffic control by pushing self-adaptation to the edge and IoT devices.

\begin{acks}
This work was supported by MITACS through the Mitacs Accelerate program and NSERC of Canada under the Discovery and Discovery Accelerator programs.
\end{acks}

\bibliographystyle{ACM-Reference-Format}
\balance
\bibliography{paper}

\linenumbers

\end{document}